\documentclass[journal]{IEEEtran}
\usepackage{amsmath,amsfonts,amssymb}
\usepackage{array}
\usepackage{caption}                    
\usepackage[caption=false]{subfig}       

\usepackage{textcomp}
\usepackage{stfloats}
\usepackage{placeins}                     
\usepackage{url}
\usepackage{xurl}                     
\usepackage{graphicx}
\usepackage{cite}
\usepackage{booktabs}
\usepackage{multirow}
\usepackage{color}
\usepackage{hyperref}
\usepackage{bm}

\AtBeginDocument{%
  \setlength{\abovedisplayskip}{4pt plus 1pt minus 1pt}
  \setlength{\belowdisplayskip}{4pt plus 1pt minus 1pt}
  \setlength{\abovedisplayshortskip}{2pt plus 1pt minus 1pt}
  \setlength{\belowdisplayshortskip}{2pt plus 1pt minus 1pt}
}

\begin{document}
\raggedbottom 

\title{From Tokens to Energy Flexibility: Quantization-Enabled Demand Response for Data Centers with LLM Inference Workloads}

\author{
        Bojun Du,~\IEEEmembership{Student Member,~IEEE},
        Xiaoyi Fan,~\IEEEmembership{Member,~IEEE},
        Ershun Du,~\IEEEmembership{Member,~IEEE}, 
        Long Chen,~\IEEEmembership{Member,~IEEE},
        Jianpei Han,
        Qingchun Hou,~\IEEEmembership{Member,~IEEE}, 
        Ning Zhang,~\IEEEmembership{Senior Member,~IEEE},
        and Chongqing Kang,~\IEEEmembership{Fellow,~IEEE}
        \thanks{Manuscript accepted November 21, 2026. This work has been accepted for publication in IEEE Transactions on Smart Grid. \copyright~2026 IEEE. Personal use of this material is permitted. Permission from IEEE must be obtained for all other uses, in any current or future media, including reprinting/republishing this material for advertising or promotional purposes, creating new collective works, for resale or redistribution to servers or lists, or reuse of any copyrighted component of this work in other works.}
        \thanks{This work is supported by the Smart-Grid National Science and Technology Major Project (2025ZD0804800).}
        \thanks{B. Du, X. Fan, E. Du, N. Zhang and C. Kang are with the Department of Electrical Engineering, Tsinghua University, Beijing, China.}
        \thanks{L. Chen is with the Department of Computing, The Hong Kong Polytechnic University, Hong Kong, China.}
        \thanks{J. Han is with the School of Energy Storage Science and Engineering, North China University of Technology, Beijing, China.}
        \thanks{Q. Hou is with Zhejiang University, Hangzhou, China.}
}

\maketitle

\markboth{IEEE Transactions on Smart Grid}{Du \emph{et al.}: Quantization-Enabled Demand Response}

\begin{abstract}
The rapid growth of large language model (LLM) inference is creating significant data-center loads that face increasing energy-management challenges under tightening grid conditions and demand response (DR) requirements.
Conventional data-center energy management mainly relies on temporal and spatial workload shifting and campus-level energy asset scheduling, but it usually treats LLM inference demand as an aggregate load.
As a result, these approaches fail to exploit the internal characteristics of LLM serving and therefore overlook the flexibility offered by LLM-specific techniques such as model quantization.
To unlock this flexibility, this paper proposes a quantization-enabled energy management framework for grid-responsive LLM inference data centers.
First, a quantization-to-power model is established to map each model--quantization configuration to a compact set of dispatchable parameters.
Second, a two-stage quantization-enabled DR model is developed to account for model instance switching, request routing, and precision selection.
Third, a multi-campus co-optimization method is introduced for DR participation by integrating grid-side electricity and carbon signals with the quantization-enabled DR model.
Case studies show that the proposed framework reduces total data-center operating cost by \textcolor{black}{31.8\%} without curtailing served token volume, validating model quantization as an effective flexibility lever for grid-responsive LLM data-center energy management.

\end{abstract}


\section{Introduction}

\IEEEPARstart{T}{he} rapid expansion of artificial intelligence (AI) data centers is driving a steep increase in electricity demand.
The International Energy Agency projects global data-center electricity consumption to more than double to roughly 945~TWh by 2030, with accelerated AI servers driving most of the growth~\cite{iea2025energyai}.
Within this surge, large language model (LLM) inference is becoming a major driver, prompting providers to build inference-oriented data centers at unprecedented scale.
For instance, OpenAI's Stargate program targets 10~GW of dedicated AI compute capacity by 2029, with over 5~GW already under development across multiple U.S. sites~\cite{openai2025stargate}.
Such fast-growing loads place substantial pressure on power supply and may aggravate grid congestion and local capacity shortages. This motivates data center operators to actively manage computing workloads, campus energy assets, and demand response (DR) flexibility to ensure reliable supply and reduce energy costs~\cite{patterson2022carbon}.

To address the supply and cost pressures, conventional data-center DR and energy-management studies mainly exploit flexibility from three perspectives: temporal shifting, spatial shifting, and campus energy management.
First, temporal flexibility is mainly achieved by deferring batch jobs~\cite{li2015batchDR,chen2013datacenterDR} or by monetizing flexible service delivery through service-level-agreement contracts~\cite{basmadjian2018greensla}.
Second, spatial flexibility shifts workloads across geographically distributed sites according to electricity prices and grid conditions~\cite{luo2015spatial,wang2026lowcarbonRL,han2026aidcScheduling}. This idea has been extended to distribution-network restoration~\cite{jian2024restorationDC}, peer-to-peer flexibility markets~\cite{jin2025p2pDC,dvorkin2025agentconcur}, and privacy-preserving coordination~\cite{liu2025hierarchicalDC}.
Third, campus energy management co-schedules IT loads with on-site renewables, storage, and backup generation~\cite{li2018idcMicrogrid,yu2018dcmicrogrid,zhang2024energyHubsDC,guo2021battery,radovanovic2023carbon}. More recent studies further couple AI data-center scheduling with distribution network operation, AI-oriented DR program design, and grid-peak power reduction in real GPU clusters~\cite{cao2023spatiotemporalISC,chen2026defer,emerald2025conductor}.
However, these approaches mainly reshape data-center demand from outside the inference service. They still treat LLM inference as aggregate power demand and do not model how requests are served inside GPU clusters. As a result, they cannot fully exploit the internal flexibility of LLM inference workloads.

\begin{figure}[!b]
\centering
\includegraphics[width=\columnwidth]{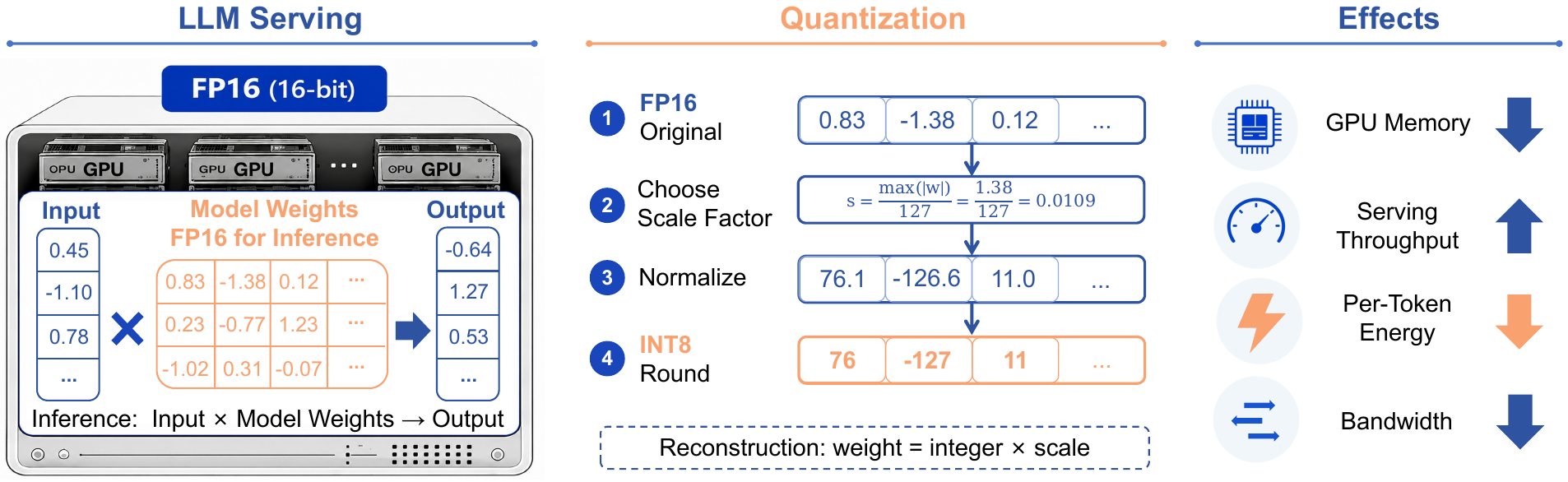}
\captionsetup{font=small}
\caption{\textcolor{black}{Illustrative FP16-to-INT8 weight quantization and its potential effects on LLM inference.}}
\label{fig:quantization_illustration}
\end{figure}

An LLM inference workload consists of requests served by resident model instances on GPU servers.
Each request occupies GPU resources, accesses model weights and runtime states in memory, and performs token-by-token numerical computation to generate the response.
Therefore, the energy consumption of LLM inference is jointly shaped by hardware power characteristics, device utilization, GPU memory occupation, and the amount of numerical computation~\cite{patel2024splitwise,niu2026tokenpowerbench,caravaca2025prompts}.
These characteristics create several levers for reducing inference power.
At the device and serving level, GPU frequency scaling can lower workload power during critical periods~\cite{wang2025loadflexibility}, parallel serving and batching improve hardware utilization~\cite{lisali2025freesh}, and matching requests with heterogeneous hardware reduces serving energy~\cite{li2025ecoserve}. 
\textcolor{black}{These methods can reduce baseline energy consumption and also provide flexibility for demand response through adjustments to hardware operation and request scheduling.}

\textcolor{black}{Beyond these device and serving control methods, adjusting the numerical precision of model instances provides an additional source of demand-response flexibility at the model-instance level. It can reduce memory occupation and computation cost without rejecting user requests.}
This precision-control mechanism offers large adjustment potential, broad applicability across hardware platforms, and fast response capability~\cite{xiao2023smoothquant,shi2025qmeter}. It is also compatible with device-level control, parallel serving and spatial workload shifting.
\textcolor{black}{Amazon Bedrock has already routed requests between models according to response-quality requirements~\cite{aws2025routingpractice}. This practice motivates exploring precision selection as an IT-side flexibility lever for grid-responsive data-center energy management.}

A primary approach to numerical-precision control is model quantization. \textcolor{black}{As illustrated in Fig.~\ref{fig:quantization_illustration}}, it represents the 16-bit weights of trained LLMs with lower-bit formats, thereby reducing memory footprint and computation cost~\cite{dettmers2022llmint8}. This coarser representation significantly lowers per-token energy consumption, while introducing a small and tunable accuracy loss~\cite{liang2026paroquant}. Thus, quantization has become a mainstream tool in the LLM community, with 8-bit and 4-bit methods serving as practical deployment options~\cite{frantar2023gptq,lin2024awq,kurtic2025give}. 

However, modeling quantization for grid-responsive data-center energy management raises three challenges. First, each model--quantization configuration must be translated into dispatchable power and serving parameters, including GPU occupancy, throughput, energy consumption, QoS degradation, and switching cost. Second, quantization-enabled flexibility couples multiple interdependent factors, including which low-precision model instances are loaded, which request types are eligible for reduced precision, and how the energy-saving benefit is balanced against QoS degradation. Third, its system-level value remains unclear in multi-campus data centers, where quantization decisions must be coordinated with workload routing, campus energy assets, electricity prices, carbon signals, and DR requirements.

\begin{figure}[!htbp]
\centering
\includegraphics[width=\columnwidth]{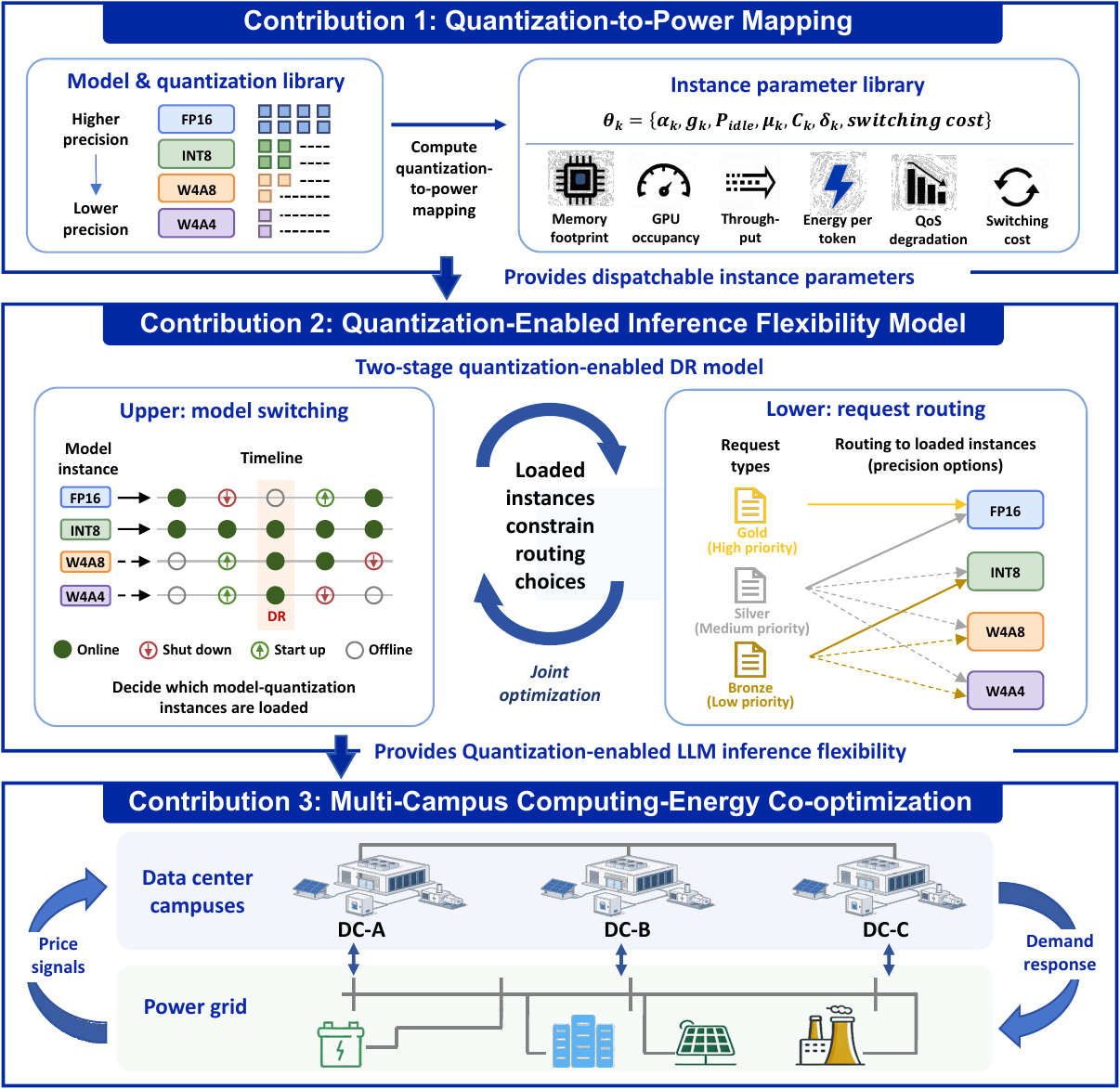}
\caption{Overview of the proposed quantization-enabled DR framework.}
\label{fig:framework_overview}
\end{figure}

To bridge these gaps, this paper models the flexibility offered by quantization, transferring this LLM inference acceleration technique into a demand-side flexibility resource for data center campuses. 
On this basis, a co-optimization method is developed for multi-campus LLM data centers, whose schematic structure is shown in Fig.~\ref{fig:framework_overview}. The main contributions are summarized as follows.
\begin{itemize}
    \item \textbf{The quantization-to-power mapping model is established for LLM inference data centers.} 
    Specifically, a parameterized modeling framework is developed to convert LLM quantization configurations into power system scheduling parameters, including instance capacity, power consumption, service quality degradation, and switching cost. This mapping lays the physical and operational groundwork for leveraging model quantization as a dispatchable demand-side flexibility resource instead of solely a computing-side compression technique.

    \item \textbf{A two-stage quantization-enabled DR model is formulated to harness LLM inference flexibility at the model-instance level.} 
    The operational flexibility of LLM inference includes two dimensions: model instance switching and request-level precision routing. The upper stage decides which model--quantization instances are loaded on GPU clusters, whereas the lower stage routes inference requests across loaded instances and chooses the precision level. This hierarchical structure effectively separates slow loading decisions from fast precision dispatch, thereby characterizing the practical flexibility boundary of LLM inference services.

    \item \textbf{A collaborative optimization method with quantization-enabled DR is proposed for multi-campus LLM data centers.} 
    Under the guidance of time-varying nodal electricity prices and carbon emission signals, this method integrates inference-side quantization flexibility into a multi-campus co-optimization framework that jointly schedules on-site gas turbines, battery storage, photovoltaic generation, and DR participation. The proposed method yields a tractable mixed-integer linear program (MILP) that aligns LLM inference with grid-side optimization, thereby reducing cost, mitigating carbon emissions, and enabling DR simultaneously.
\end{itemize}

The remainder of this paper is organized as follows. 
Section~\ref{sec:quantization_power_model} develops the quantization-to-power modeling framework that converts each model--quantization configuration into a set of dispatchable scheduling parameters. 
Section~\ref{sec:two_stage_dr} formulates the two-stage quantization-enabled demand response model that couples model instance switching with request-level precision routing. 
Section~\ref{sec:campus_grid_model} embeds the inference-side flexibility in a multi-campus energy co-optimization with electricity price and carbon signals. 
Section~\ref{sec:case_study} reports the case studies on a representative multi-campus system, and Section~\ref{sec:conclusion} concludes the paper.

\section{Quantization-to-Power Mapping for LLM Inference}
\label{sec:quantization_power_model}

LLM quantization affects data-center power consumption because it changes how each loaded model instance occupies GPU memory and serves inference requests. In LLM serving, model weights, activations, and runtime states are stored or accessed in GPU memory, while output tokens are generated through repeated numerical computation~\cite{kwon2023vllm}. Quantization reduces the numerical precision of these tensors from conventional 16-bit formats to lower bit-width representations, such as 8-bit or 4-bit formats~\cite{dettmers2022llmint8,lin2024awq}. Since LLM inference is often constrained by memory movement, batching state, and serving throughput~\cite{patel2024asplos,niu2026tokenpowerbench}, lowering precision can reduce memory footprint, increase instance throughput, and decrease dynamic energy per token. Recent quantization-system measurements further show that these efficiency gains come with method- and task-dependent quality degradation~\cite{shi2025qmeter}. Therefore, quantization provides a controllable quality--power tradeoff that can be exploited for demand response.

This paper studies quantization from a data-center energy scheduling perspective rather than modeling millisecond-level GPU serving dynamics.
Metrics such as time to first token and inter-token latency are important for serving-system design, but they evolve at a much finer timescale than 15-min DR scheduling.
Accordingly, the proposed model operates at the grid-responsive energy-scheduling layer and represents LLM serving using token throughput and per-token energy.
\textcolor{black}{These parameters define the capacity envelope under a reference workload and deployment configuration for 15-min resource scheduling, rather than describing instantaneous request-level resource usage. This parameterization is consistent with Microsoft's capacity-sizing
practice of using representative workload statistics, including average input and output lengths, to estimate deployment capacity \cite{Microsoft2026PTUSizing}.}

\subsection{Instance Parameterization}

Under this timescale separation, this section does not develop a new quantization algorithm or a cycle-level GPU simulator. Instead, it establishes a compact mapping from each model--quantization configuration to the scheduling parameters required by the power-system optimization model. For each model family \(m\in\mathcal{M}\) and quantization configuration \(q\in\mathcal{Q}_m\), an inference instance type is defined as
\begin{equation}
    k \equiv (m,q), \quad k\in\mathcal{K}.
\end{equation}
The mapping outputs the parameter set
\begin{equation}
    \Theta_k =
    \left\{
    a_k,\,
    g_k,\,
    p^{\rm idle}_k,\,
    \mu_k,\,
    e_k,\,
    \delta_k,\,
    \tau^{\ell}_k,\,
    \tau^{u}_k,\,
    E^{\ell}_k,\,
    E^{u}_k
    \right\},
    \label{eq:instance_parameter_set}
\end{equation}
where $a_k$ and $g_k$ denote the GPU count and total GPU memory occupied by one loaded instance, $p^{\rm idle}_k$ is the idle power, $\mu_k$ is the maximum token throughput, 
\textcolor{black}{$e_k$ is the equivalent active energy per output token and $\delta_k$ is the QoS degradation relative to the FP16 baseline.} 
The parameters $\tau^{\ell}_k$ and $\tau^{u}_k$ denote the loading and unloading time of instance $k$, respectively, while $E^{\ell}_k$ and $E^{u}_k$ denote the corresponding loading and unloading energy. Thus, $\Theta_k$ acts as the interface that converts precision control into dispatchable power-system quantities.

\textcolor{black}{Among these parameters, $e_k$ relies on one additional aggregation assumption. At the 15-min dispatch resolution, we assume that each interval contains a sufficiently large number of requests for the interval-level input-to-output token ratio to vary only slightly.
This aggregate representation is motivated by production-trace studies that characterize LLM workloads in terms of request-arrival processes and input/output-length distributions \cite{wang2025burstgpt,xiang2026servegen}. 
Accordingly, the average prefill contribution is incorporated into $e_k$, while output tokens serve only as the common unit for workload accounting. The empirical validity and applicability boundary of this approximation are examined in Appendix~\ref{app:equivalent_energy_validation}.}

\subsection{Resource Footprint and GPU Occupancy}

For a transformer model \(m\) with \(P_m\) parameters, the effective weight bit-width of quantization configuration \(q\) includes both quantized weights and group-wise scale/zero-point metadata~\cite{frantar2023gptq,lin2024awq}:
\begin{equation}
    \beta_q = b^{w}_q + \frac{b^{s}_q+b^{z}_q}{G_q},
    \label{eq:effective_bitwidth}
\end{equation}
where \(b^{w}_q\) is the weight bit-width, \(b^{s}_q\) and \(b^{z}_q\) are metadata bit-widths, and \(G_q\) is the quantization group size. The corresponding weight footprint is \(M^w_{m,q}=P_m\beta_q/8\).

The runtime memory also includes the KV cache, activation buffers, and serving-framework overhead. 
\textcolor{black}{For scheduling-level capacity estimation, $B$ denotes the reference batch size and $S$ denotes the reference cached context length per sequence. The corresponding KV-cache requirement is estimated as}
\begin{equation}
    M^{\rm KV}_{m,q}
    =
    2L_m h^{\rm KV}_m d^h_m B S \cdot \frac{b^{\rm KV}_q}{8},
    \label{eq:kv_memory}
\end{equation}
where \(L_m\), \(h^{\rm KV}_m\), and \(d^h_m\) are the layer count, KV head count, and head dimension. The factor \(2\) accounts for keys and values.
\textcolor{black}{The product $BS$ represents the reference total cached-token count used to construct the capacity envelope. 
It does not imply that all requests have identical lengths or that the actual KV-cache occupancy remains constant during inference.}
\textcolor{black}{
For scheduling-level resource estimation, we treat the mean output length as approximately unchanged across quantization configurations. 
This approximation is motivated by reported cases of limited output-length variation~\cite{liu2025quantizationreasoning,marcuzzi2026quantizationbias},
with complementary experimental analysis provided in Appendix \ref{app:quantization_length_subset}.
}

The total memory footprint used by the dispatch model is
\begin{equation}
    M^{\rm tot}_{m,q}
    =
    M^w_{m,q}
    +
    M^{\rm KV}_{m,q}
    +
    M^{\rm act}_{m,q}
    +
    M^{\rm ovh}_{m,q},
    \label{eq:total_memory}
\end{equation}
where \(M^{\rm act}_{m,q}\) denotes activation and runtime-buffer memory, and \(M^{\rm ovh}_{m,q}\) captures framework overhead and reserved GPU memory. Given the per-GPU memory \(G^{\rm GPU}\), the scheduling parameters are
\begin{equation}
    g_k = \frac{M^{\rm tot}_{m,q}}{2^{30}},
    \quad
    a_k = \left\lceil \frac{g_k}{G^{\rm GPU}} \right\rceil,
    \label{eq:gpu_requirement}
\end{equation}
\textcolor{black}{Thus, $g_k$ represents the per-instance GPU memory requirement for capacity allocation under the reference operating condition, rather than an instantaneous memory reading.}
Reducing precision therefore changes both the continuous memory footprint \(g_k\) and the integer GPU occupancy \(a_k\) of a loaded instance.

\subsection{Throughput, Energy, and Quality Parameters}
\label{sec:throughput_energy_quality}

Besides resource occupancy, each instance is characterized by serving throughput, token energy, and quality. 
\textcolor{black}{The parameter $\mu_k$ denotes the maximum output-token throughput of instance type $k$ that satisfies the specified latency requirements under a given workload and serving configuration.} 
Since practical LLM throughput depends on batching policy, serving runtime, parallelization strategy, and request mix, $\mu_k$ is treated as a calibrated parameter obtained from existing LLM-serving measurements and public benchmarks~\cite{williams2009roofline,llminferencebench2024,nvidia2023tensorrtllm}.
\textcolor{black}{Together with the resource parameters, it is held fixed for scheduling-level capacity estimation under given deployment conditions. 
Appendix~\ref{app:equivalent_energy_validation} provides an empirical assessment of interval-level average throughput stability.}

\textcolor{black}{Following the interval-level definition above, we estimate the equivalent active-energy coefficient \(e_k\) from the average computation, memory-access, and communication costs as}
\begin{equation}
    e_k =
    \eta^E_k
    \left(
    \epsilon^{\rm op}_q c^{\rm op}_{m,q}
    +
    \epsilon^{\rm mem} c^{\rm mem}_{m,q}
    +
    \epsilon^{\rm comm} c^{\rm comm}_{m,q}
    \right),
    \label{eq:dynamic_energy}
\end{equation} 
where \(c^{\rm op}_{m,q}\), \(c^{\rm mem}_{m,q}\), and \(c^{\rm comm}_{m,q}\) represent per-token computation, memory-traffic, and communication requirements, while \(\epsilon^{\rm op}_q\), \(\epsilon^{\rm mem}\), and \(\epsilon^{\rm comm}\) are the corresponding unit energy costs.
The factor \(\eta^E_k\) calibrates serving-stack effects such as kernel efficiency, batching, and parallelization overheads. This decomposition follows hardware energy models and token-level LLM power measurements~\cite{patel2024asplos,niu2026tokenpowerbench}.

The idle power of a loaded instance is modeled separately as
\begin{equation}
    p^{\rm idle}_k
    =
    a_k p^{\rm idle,GPU} + p^{\rm host}_k ,
    \label{eq:idle_power}
\end{equation}
where \(p^{\rm idle,GPU}\) is per-GPU resident idle power and \(p^{\rm host}_k\) is the allocated host-side overhead. Since \(e_k\) denotes only activity-dependent token energy, separating \(p^{\rm idle}_k\) avoids double counting idle energy~\cite{patel2024asplos,wattcounts2026,niu2026tokenpowerbench}.

Finally, quantization quality is represented by a normalized degradation factor:
\begin{equation}
    \delta_{m,q}
    =
    \max_{r\in\mathcal{R}_m}
    \frac{S_{m,{\rm FP16},r}-S_{m,q,r}}{S_{m,{\rm FP16},r}},
    \label{eq:qos_degradation}
\end{equation}
where \(S_{m,q,r}\) is the normalized score of model \(m\) under configuration \(q\) on metric \(r\), and \(\mathcal{R}_m\) is the set of evaluation metrics or tasks used for model \(m\). The FP16 baseline has \(\delta_{m,q}=0\), and lower-is-better metrics such as perplexity are converted before normalization. For instance type \(k=(m,q)\), we set \(\delta_k=\delta_{m,q}\). The values used in the case study are calibrated from reported quantization evaluations~\cite{dettmers2022llmint8,frantar2023gptq,shao2024omniquant,shi2025qmeter}.

\subsection{Switching Dynamics and Power-System Coupling}
\label{sec:bridge}

Model switching is not instantaneous. 
When a model instance is loaded, the checkpoint must be transferred from storage to GPU memory, the serving runtime must initialize the instance, and the corresponding execution context must be prepared.
When an instance is unloaded, its occupied memory and runtime resources are released.
Following LLM serving measurements~\cite{fu2024serverlessllm}, the loading and unloading times, denoted by $\tau_k^\ell$ and $\tau_k^u$, are treated as calibrated parameters for each instance type $k$.
The corresponding switching energy terms are
\begin{align}
    E_k^\ell
    &=
    p_k^\ell\tau_k^\ell,
    \label{eq:switching_energy_model}\\
    E_k^u
    &=
    p_k^u \tau_k^u ,
\end{align}
where \(p_k^\ell\) and \(p_k^u\) are the loading and unloading powers.

Combining the above parameters, the IT power of campus \(n\) at period \(t\) is
\begin{equation}
\begin{aligned}
    P^{\rm IT}_{n,t} = {\rm PUE}_n
    \bigg[
    &\sum_{k}p^{\rm idle}_k x_{k,n,t}
    + \sum_{k}\sum_{j} e_k \lambda_{k,n,j,t} \\
    &+ \frac{1}{\Delta \tau^{s}}\sum_{k}
    \left( E^{\ell}_k
    y_{k,n,t+1}
    + E^{u}_k z_{k,n,t} \right)
    \bigg],
\end{aligned}
\label{eq:it_power}
\end{equation}
\textcolor{black}{where $x_{k,n,t}$ counts ready instances and $\lambda_{k,n,j,t}$ is the routed token rate. The variables $y_{k,n,t}$ and $z_{k,n,t}$ count instances becoming ready and beginning unloading at the start of interval $t$, respectively. 
Loading $y_{k,n,t+1}$ occurs in interval $t$, so $E_k^\ell$ includes their full loading energy.}
The three terms in \eqref{eq:it_power} represent resident idle power, activity-dependent inference power, and switching power.

Equation~\eqref{eq:it_power} is the key bridge from LLM quantization to power-system scheduling. Quantization affects grid-facing IT demand through three channels: instance switching, governed by \((a_k,g_k,p^{\rm idle}_k)\); request routing and precision selection, governed by \((\mu_k,e_k,\delta_k)\); and switching dynamics, governed by \((\tau^\ell_k,\tau^u_k,E^\ell_k,E^u_k)\). The two-stage DR model in Section~\ref{sec:two_stage_dr} exploits these channels to coordinate precision control with model loading and campus energy operation.

\section{Two-Stage Quantization-Enabled Demand Response Model}
\label{sec:two_stage_dr}

\subsection{Overview of the Two-Stage Decision Hierarchy}

Based on the quantization-to-power mapping derived in Section~\ref{sec:quantization_power_model}, this section formulates the IT-side flexibility of LLM inference data centers, which is modeled as a two-stage decision process. The first stage determines which model--quantization instances are loaded on GPU clusters. This stage is constrained by GPU resources, memory capacity, and model loading/unloading dynamics, and therefore changes at the energy scheduling timescale. The second stage routes inference requests to loaded instances and selects the precision level used for service. The two stages are solved jointly as a single MILP. The coupling between them is established by the instance capacity constraint and the IT power equation. Specifically, the switching variables determine the available serving capacity, while the routing variables determine the dynamic inference power and QoS degradation.

\subsection{Stage 1: Model Instance Switching}

\textcolor{black}{We assume that outgoing instances stop accepting new requests sufficiently early to complete all in-flight requests on the original instances before the interval boundary. The lower-level serving controller verifies that these instances are drained before initiating physical unloading~\cite{microsoft2026modelmigration}.}
\textcolor{black}{For intervals $t=1,\ldots,T$, $x_{k,n,t}$ counts ready instances of type $k$ at data center $n$. Incoming instances $y_{k,n,t}$ load during the final $\tau_k^\ell$ seconds of interval $t-1$ and provide no service before interval $t$. Outgoing instances $z_{k,n,t}$ provide no service in interval $t$ and release their resources $\tau_k^u$ seconds after its start.}
Thus, the instance state transition is
\textcolor{black}{
\begin{equation}
\begin{array}{l}
x_{k,n,t}-x_{k,n,t-1}=y_{k,n,t}-z_{k,n,t}, \\
z_{k,n,t}\le x_{k,n,t-1},
\quad \forall k,n,\ t=2,\ldots,T.
\end{array}
\label{eq:instance_transition}
\end{equation}}

To avoid simultaneous loading and unloading of the same instance type, a binary variable \mbox{$b_{k,n,t}$} is introduced:
\begin{align}
y_{k,n,t} &\leq M^{\max}_{k,n} b_{k,n,t}, \label{eq:start_limit}\\
z_{k,n,t} &\leq M^{\max}_{k,n} (1-b_{k,n,t}),
\label{eq:stop_limit}
\end{align}
where $M^{\max}_{k,n}$ is an upper bound on the number of instances of type $k$ that can be hosted by data center $n$, given by
\begin{equation}
M^{\max}_{k,n}=\left\lfloor \frac{A_n}{a_k}\right\rfloor ,
\label{eq:max_instance}
\end{equation}
where $A_n$ is the total number of GPUs in data center $n$, and $a_k$ is the number of GPUs required by one loaded instance of type $k$. 

{\color{black}
GPU and memory consumption must include not only serving instances $x_{k,n,t}$ but also instances loading for the next interval $y_{k,n,t+1}$ and those unloading in the current interval $z_{k,n,t}$. 
Let $a_k$ and $g_k$ denote the GPU count and GPU memory required by one instance of type $k$, and let $A_n$ and $G_n$ denote the corresponding capacities of data center $n$. The resource constraints are
\begin{align}
\sum_{k\in\mathcal K} a_k x_{k,n,t}
+\sum_{k\in\mathcal L_h} a_k y_{k,n,t+1}
+\sum_{k\in\mathcal U_h} a_k z_{k,n,t}
&\le A_n,
\label{eq:gpu_count_limit}\\
\sum_{k\in\mathcal K} g_k x_{k,n,t}
+\sum_{k\in\mathcal L_h} g_k y_{k,n,t+1}
+\sum_{k\in\mathcal U_h} g_k z_{k,n,t}
&\le G_n,
\label{eq:gpu_memory_limit}
\end{align}
where the three terms represent resources occupied by serving, loading, and unloading
instances, respectively.
GPUs released after unloading may be reused for loading later in the same interval. To capture this reuse, we define a set of occupancy patterns $\mathcal H$. Each pattern $h$ specifies
two model-type sets: $\mathcal L_h$ for instances currently loading and $\mathcal U_h$ for instances still unloading. These sets are determined before optimization by ordering the fixed loading-start and unloading-completion times. 
Enforcing the constraints for every pattern limits peak resource occupancy without introducing finer-resolution scheduling decisions.
For example, Fig.~\ref{fig:switching_gpu_occupancy} illustrates three
patterns:
$(\mathcal U_1,\mathcal L_1)=(\{A\},\{C\})$,
$(\mathcal U_2,\mathcal L_2)=(\varnothing,\{C\})$, and
$(\mathcal U_3,\mathcal L_3)=(\varnothing,\{B,C\})$.
GPUs released by A can support B's subsequent loading, whereas C requires separate resources because its loading overlaps with A's unloading. This formulation reserves resources for overlapping transitions while allowing reuse between non-overlapping operations.

\begin{figure}[!htbp]
    \centering
    \includegraphics[width=\columnwidth]{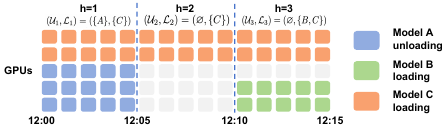}
    \caption{\textcolor{black}{
    GPU occupancy and reuse under the prescribed transition timing.
    Gray blocks indicate resources available for reuse.
    Continuously serving instances are omitted.
    }}
    \label{fig:switching_gpu_occupancy}
\end{figure}
}

\subsection{Stage 2: Request Routing and Precision Selection}

Inference demand is classified by request type \(j\in\mathcal{J}\), where each request type corresponds to a specific model family and QoS tier. Let \(D_{j,n,t}\) be the demand of request type \(j\) assigned to data center \(n\), and $\lambda_{k,n,j,t}$ be the serving rate of request type $j$ routed to instance type $k$. The global demand balance is
\begin{equation}
\sum_{n\in\mathcal{N}}D_{j,n,t}=D^{\rm total}_{j,t},
\quad \forall j,t .
\label{eq:global_demand_balance}
\end{equation}

Within each data center, the allocated demand must be fully served by instances of the required model family:
\begin{equation}
\sum_{k\in\mathcal{K}_j}\lambda_{k,n,j,t}=D_{j,n,t},
\quad \forall n,j,t ,
\label{eq:local_dispatch_balance}
\end{equation}
where \(\mathcal{K}_j\) is the set of instance types with the same model family as request type \(j\).

\textcolor{black}{The routed requests cannot exceed the available online service capacity:}
{\color{black}
\begin{equation}
\begin{aligned}
\sum_{j\in\mathcal{J}}\frac{\lambda_{k,n,j,t}}{\mu_k}
\leq {} & \rho x_{k,n,t}, \quad \forall k,n,t,
\end{aligned}
\label{eq:instance_capacity}
\end{equation}
}
\textcolor{black}{where $0<\rho\leq 1$ is a capacity redundancy factor to represent the capacity margin reserved for short-term request fluctuations and forecast errors.}
\textcolor{black}{Only ready instances contribute to service capacity. Incoming instances finish loading at the interval boundary, so no loading-time deduction is applied to $x_{k,n,t}$. Outgoing instances are already excluded by \eqref{eq:instance_transition} and provide no service during unloading.}

For incompatible instance--request pairs, routing is prohibited:
\begin{equation}
\lambda_{k,n,j,t}=0,
\quad \forall k\notin\mathcal{K}_j,\;n,t .
\label{eq:incompatible_routing}
\end{equation}

Equations~\eqref{eq:local_dispatch_balance}--\eqref{eq:incompatible_routing} enable the optimizer to select both the serving location and the quantization level. For example, during DR events, low-priority requests may be routed to GPTQ or OmniQuant instances if the resulting QoS degradation is acceptable.

\subsection{QoS Degradation and Service Compatibility}

The model-family compatibility constraints above determine which configurations can serve each request type. Let \(\delta_k\) denote the calibrated degradation indicator of model--quantization configuration \(k\), and let \(\delta_j^{\max}\) denote the maximum configuration degradation allowed for request type \(j\). 
To prevent higher-quality traffic from compensating for assignments to configurations with excessive degradation, we impose the following quality eligibility constraint:
{\color{black}
\begin{equation}
(\delta_k-\delta_j^{\max})\lambda_{k,n,j,t}\leq 0,
\quad \forall k\in\mathcal{K}_j,\;n,j,t .
\label{eq:qos_endpoint_eligibility}
\end{equation}}
\textcolor{black}{Since \(\lambda_{k,n,j,t}\geq0\), any configuration with \(\delta_k>\delta_j^{\max}\) must receive zero service allocation for request type \(j\). 
Thus, requests can only be routed to configurations that satisfy their service-tier quality threshold.}

The QoS penalty at data center \(n\) and period \(t\) is
\begin{equation}
Q_{n,t}
=
\Delta\tau^{s}
\sum_{j\in\mathcal{J}}
C_j^{q}
\sum_{k\in\mathcal{K}_j}
\delta_k\lambda_{k,n,j,t},
\label{eq:qos_penalty}
\end{equation}
where \(C_j^{q}\) is the penalty coefficient of request type \(j\), expressed in cost per token.

Together, \eqref{eq:instance_transition}--\eqref{eq:qos_penalty} define the feasible IT-side operating set of the LLM inference cluster. For any feasible solution, the corresponding IT power is calculated by \eqref{eq:it_power}, which aggregates resident idle power, routing-dependent inference power, and switching energy. This power becomes the IT load input to the campus energy co-optimization model in Section~\ref{sec:campus_grid_model}.

\section{Multi-Campus Energy Co-Optimization with Electricity-Carbon Signals}
\label{sec:campus_grid_model}

This section embeds the LLM inference flexibility model into a multi-campus energy co-optimization problem. Each data center campus is connected to a distribution network and equipped with behind-the-meter energy assets. The grid side provides nodal electricity prices and carbon-intensity signals, while the campus side optimizes LLM inference, on-site generation, battery operation, renewable utilization, and DR participation.

\subsection{Multi-Campus System Architecture}

Consider a set of geographically distributed AI data center campuses \(\mathcal{N}\). Each campus \(n\in\mathcal{N}\) contains a GPU inference cluster, an on-site gas turbine, a battery energy storage system (BESS), photovoltaic (PV) generation, and a grid connection. The GPU cluster is described by the two-stage model in Section~\ref{sec:two_stage_dr}. The remaining energy assets are scheduled at the same 15-min dispatch interval.

The campus-level power balance is
\begin{equation}
P^{\rm grid}_{n,t}
+
P^{\rm gen}_{n,t}
+
P^{\rm dis}_{n,t}
+
P^{\rm pv}_{n,t}
=
P^{\rm IT}_{n,t}
+
P^{\rm ch}_{n,t},
\quad \forall n,t .
\label{eq:campus_power_balance}
\end{equation}
where $P^{\rm grid}_{n,t}$ is the power purchased from the distribution grid, $P^{\rm gen}_{n,t}$ is the on-site gas-turbine output, $P^{\rm dis}_{n,t}$ is the BESS discharging power, and $P^{\rm pv}_{n,t}$ is the utilized PV generation. On the demand side, $P^{\rm IT}_{n,t}$ denotes the IT power of the LLM inference workload, and $P^{\rm ch}_{n,t}$ denotes the BESS charging power. All variables are defined for campus $n$ and time period $t$. This equation is the physical coupling between quantization-enabled IT flexibility and campus-level energy operation.

\subsection{Campus Energy Asset Models}

\subsubsection{Gas Turbine}

Let \(u^{\rm gen}_{n,t}\), \(v^{\rm gen}_{n,t}\), and \(w^{\rm gen}_{n,t}\) denote the on/off, startup, and shutdown status of the gas turbine. The commitment transition is
\begin{equation}
u^{\rm gen}_{n,t}-u^{\rm gen}_{n,t-1}
=
v^{\rm gen}_{n,t}-w^{\rm gen}_{n,t},
\quad \forall n,t .
\label{eq:gt_transition}
\end{equation}

The generation output is bounded by
\begin{equation}
\underline{P}^{\rm gen}_{n}u^{\rm gen}_{n,t}
\leq
P^{\rm gen}_{n,t}
\leq
\overline{P}^{\rm gen}_{n}u^{\rm gen}_{n,t},
\quad \forall n,t .
\label{eq:gt_output_bound}
\end{equation}
where $\underline{P}^{\rm gen}_{n}$ and $\overline{P}^{\rm gen}_{n}$ are the minimum and maximum generation limits. 

Ramp limits are modeled as
\begin{align}
P^{\rm gen}_{n,t}-P^{\rm gen}_{n,t-1}
&\leq
R^{\rm up}_{n}u^{\rm gen}_{n,t-1}
+
R^{\rm su}_{n}v^{\rm gen}_{n,t},
\label{eq:gt_ramp_up}\\
P^{\rm gen}_{n,t-1}-P^{\rm gen}_{n,t}
&\leq
R^{\rm dn}_{n}u^{\rm gen}_{n,t}
+
R^{\rm sd}_{n}w^{\rm gen}_{n,t}.
\label{eq:gt_ramp_down}
\end{align}
where $R^{\rm up}_{n}$ and $R^{\rm dn}_{n}$ are the ramp-up and ramp-down limits. $R^{\rm su}_{n}$ and $R^{\rm sd}_{n}$ are the startup and shutdown ramp allowances. 

The minimum up/down time constraints are
\begin{align}
\sum_{\tau=t}^{t+T^{\rm on}_{n}-1}
u^{\rm gen}_{n,\tau}
&\geq
T^{\rm on}_{n}v^{\rm gen}_{n,t},
\quad \forall n,t,
\label{eq:gt_min_up}\\
\sum_{\tau=t}^{t+T^{\rm off}_{n}-1}
(1-u^{\rm gen}_{n,\tau})
&\geq
T^{\rm off}_{n}w^{\rm gen}_{n,t},
\quad \forall n,t.
\label{eq:gt_min_down}
\end{align}
where $T^{\rm on}_{n}$ and $T^{\rm off}_{n}$ are the minimum on-time and off-time of the turbine. 

\subsubsection{Battery Energy Storage}

Let the binary variables $u^{\rm ch}_{n,t}$ and $u^{\rm dis}_{n,t}$ indicate whether the BESS is charging or discharging. The charge and discharge powers are limited by
\begin{align}
0 \leq P^{\rm ch}_{n,t}
&\leq
\overline{P}^{\rm ch}_{n}u^{\rm ch}_{n,t},
\quad \forall n,t,
\label{eq:bess_ch_limit}\\
0 \leq P^{\rm dis}_{n,t}
&\leq
\overline{P}^{\rm dis}_{n}u^{\rm dis}_{n,t},
\quad \forall n,t,
\label{eq:bess_dis_limit}\\
u^{\rm ch}_{n,t}+u^{\rm dis}_{n,t}
&\leq 1,
\quad \forall n,t.
\label{eq:bess_mutex}
\end{align}
where $\overline{P}^{\rm ch}_{n}$ and $\overline{P}^{\rm dis}_{n}$ are the maximum charging and discharging powers. 

The state of energy evolves as
\begin{equation}
E^{\rm bat}_{n,t+1}
=
E^{\rm bat}_{n,t}
+
\left(
\eta^{\rm ch}_{n}P^{\rm ch}_{n,t}
-
\frac{P^{\rm dis}_{n,t}}{\eta^{\rm dis}_{n}}
\right)\Delta\tau,
\quad \forall n,t,
\label{eq:bess_energy_balance}
\end{equation}
where $E^{\rm bat}_{n,t}$ is the stored energy, $\eta^{\rm ch}_{n}$ and $\eta^{\rm dis}_{n}$ are the charging and discharging efficiencies, and $\Delta\tau$ is the dispatch interval in hours. The stored energy is bounded by
\begin{equation}
\underline{E}^{\rm bat}_{n}
\leq
E^{\rm bat}_{n,t}
\leq
\overline{E}^{\rm bat}_{n}, \ E^{\rm bat}_{n,0}=E^{\rm bat,0}_{n}, \ E^{\rm bat}_{n,T}\geq E^{\rm bat,0}_{n}
\label{eq:bess_energy_bound}
\end{equation}
where $\underline{E}^{\rm bat}_{n}$ and $\overline{E}^{\rm bat}_{n}$ are the minimum and maximum energy limits, and $E^{\rm bat,0}_{n}$ is the initial state. 
A terminal constraint is imposed to avoid end-of-horizon depletion.

\subsubsection{Photovoltaic Generation}

PV utilization and curtailment satisfy
\begin{equation}
P^{\rm pv}_{n,t}+P^{\rm curt}_{n,t}
=
\alpha^{\rm pv}_{n,t}\overline{P}^{\rm pv}_{n},
\quad
P^{\rm pv}_{n,t},P^{\rm curt}_{n,t}\geq 0,
\quad \forall n,t .
\label{eq:pv_model}
\end{equation}
where $P^{\rm pv}_{n,t}$ is the utilized PV power, $P^{\rm curt}_{n,t}$ is the curtailed PV power, $\overline{P}^{\rm pv}_{n}$ is the installed PV capacity, and $\alpha^{\rm pv}_{n,t}$ is the PV availability factor.

\subsection{Demand Response Baseline, Commitment, and Shortfall}

\textcolor{black}{Let \(P^{\rm base}_{n,t}\) denote the baseline grid purchase of campus \(n\) at time \(t\). For each operating strategy, this baseline is obtained by optimizing normal operation without DR participation, while retaining the same scheduling capabilities, workload and service-quality requirements. The resulting trajectory is then fixed when optimizing DR participation.}
During the DR window \(\mathcal{T}_{\rm dr}\), the committed DR capacity \(R_{n,t}\) and shortfall \(S_{n,t}\) satisfy
\begin{align}
P^{\rm base}_{n,t}
-
P^{\rm grid}_{n,t}
+
S_{n,t}
&\geq
R_{n,t},
\quad \forall n,t\in\mathcal{T}_{\rm dr},
\label{eq:dr_delivery}\\
0\leq R_{n,t}
\leq
\overline{R}_{n,t},
\quad
& S_{n,t}\geq 0,
\quad \forall n,t\in\mathcal{T}_{\rm dr}.
\label{eq:dr_bound}
\end{align}
For \(t\notin\mathcal{T}_{\rm dr}\), \(R_{n,t}=S_{n,t}=0\).
The DR performance is measured against grid purchase \(P^{\rm grid}_{n,t}\), rather than total IT load. Therefore, the campus can deliver DR through three coordinated channels: IT power reduction by quantization-enabled routing, battery discharge, and increased on-site generation.


\subsection{Distribution-Level Price and Carbon-Intensity Signals}

\textcolor{black}{The distribution grid and data center campuses are operated by different entities, so it is difficult to merge their decisions into one centralized optimization problem in practice.
Therefore, the grid operator first solves the distribution-level operation model and sends nodal electricity prices $\pi^{e}_{n,t}$ and carbon intensities $\pi^{c}_{n,t}$ to the data center operator. The data center operator then treats these signals as parameters and optimizes inference scheduling, local energy assets, and DR participation.
Similar signal-based approaches have been used in carbon-aware data center scheduling and low-carbon demand response, where users adjust their operation in response to externally supplied carbon information~\cite{radovanovic2023carbon,li2022mechanism,eso2024carbon}.}

\subsubsection{SOCP-ACOPF for Nodal Price}

The nodal electricity price is calculated from a per-period AC optimal power flow (ACOPF) model of the distribution feeder.
For each time period, the ACOPF minimizes the grid operating cost subject to power-flow, voltage, and line-capacity constraints.
The price $\pi^{e}_{n,t}$ is then obtained as the marginal cost of serving an additional active load at the bus connected to campus $n$.
This procedure captures both temporal price variation driven by system operating conditions and spatial price variation caused by network congestion and losses.

\subsubsection{Carbon Emission Flow for Nodal Carbon Intensity}

The carbon emission flow model traces carbon responsibility from generation sources to users. Based on active power flows, it enables nodal-level carbon-emission tracking of electricity consumption at a 15-min resolution~\cite{kang2012cef}. Let \(P_{B,t}\) be the branch power-flow distribution matrix, \(P_{G,t}\) the generator injection matrix, and \(E_{G,t}\) the vector of generator emission rates. The nodal carbon intensity vector is
\begin{equation}
E_{N,t}
=
(P_{N,t}-P_{B,t}^{\top})^{-1}
P_{G,t}^{\top}
E_{G,t},
\label{eq:cef_model}
\end{equation}
where \(P_{N,t}\) is the diagonal matrix of nodal through-power, and \(i(n)\) maps campus \(n\) to its connected distribution-network bus. The carbon intensity faced by campus \(n\) is
\begin{equation}
\pi^{c}_{n,t}=E_{N,t}^{i(n)}.
\label{eq:campus_carbon_intensity}
\end{equation}
This dynamic carbon signal captures both temporal and spatial variation caused by network power flows.

\subsection{Integrated MILP Objective and Complete Formulation}

The multi-campus data center operator minimizes its total operating cost, which consists of grid electricity purchase cost, carbon cost, on-site turbine fuel and start--stop costs, BESS cycling cost, QoS degradation cost, and DR reward/shortfall settlement. 
Thus, the objective is
\begin{equation}
\begin{aligned}
\min \quad
&
\sum_{n,t}
(\pi^{e}_{n,t}+\kappa^{c}\pi^{c}_{n,t})
P^{\rm grid}_{n,t}\Delta\tau
\\
&+
\sum_{n,t}
\left(
\frac{c^{\rm fuel}_{n}}{\eta^{e}_{n}}
+
\frac{\kappa^{c}e^{\rm fuel}_{n}}{\eta^{e}_{n}}
\right)
P^{\rm gen}_{n,t}\Delta\tau
\\
&+
\sum_{n,t}
c^{\rm cyc}_{n}
(P^{\rm ch}_{n,t}+P^{\rm dis}_{n,t})\Delta\tau
\\
&+
\sum_{n,t}
\left(
C^{\rm su,g}_{n}v^{\rm gen}_{n,t}
+
C^{\rm sd,g}_{n}w^{\rm gen}_{n,t}
\right)+
\sum_{n,t}Q_{n,t}
\\
&-
\sum_{n}\sum_{t\in\mathcal{T}_{\rm dr}}
\pi^{\rm dr}R_{n,t}\Delta\tau
+
\sum_{n}\sum_{t\in\mathcal{T}_{\rm dr}}
\pi^{\rm pen}S_{n,t}\Delta\tau .
\end{aligned}
\label{eq:integrated_objective}
\end{equation}

The complete optimization problem is
\begin{equation}
\begin{aligned}
\min \quad
& \eqref{eq:integrated_objective} \\
\text{s.t.}\quad
& \eqref{eq:instance_transition}-\eqref{eq:gpu_memory_limit},
\qquad \text{instance switching},\\
& \eqref{eq:global_demand_balance}-\eqref{eq:qos_penalty},
\qquad \text{request routing and QoS},\\
& \eqref{eq:it_power},
\qquad \ \ \ \ \ \ \ \  \text{IT power coupling},\\
& \eqref{eq:campus_power_balance}-\eqref{eq:pv_model},
\qquad \text{campus energy assets},\\
& \eqref{eq:dr_delivery}-\eqref{eq:dr_bound},
\qquad \text{DR participation},\\
& x_{k,n,t},y_{k,n,t},z_{k,n,t}\in\mathbb{Z}_{+},
b_{k,n,t}\in\{0,1\},
\forall k,n,t,\\
& u^{\rm gen}_{n,t},v^{\rm gen}_{n,t},w^{\rm gen}_{n,t},
u^{\rm ch}_{n,t},u^{\rm dis}_{n,t}\in\{0,1\},
 \forall n,t .
\end{aligned}
\label{eq:complete_milp}
\end{equation}

Since \(\pi^{e}_{n,t}\) and \(\pi^{c}_{n,t}\) are precomputed from the grid-side ACOPF and carbon emission flow models, all remaining nonlinearities are avoided. Therefore, \eqref{eq:complete_milp} is a mixed-integer linear program that jointly optimizes quantization-enabled inference operation, campus energy scheduling, and DR participation across multiple data center campuses.

\section{Case Study}
\label{sec:case_study}

\subsection{Experimental Setup}
\label{sec:setup}

\textbf{System.} Three LLM data center campuses are connected to buses 18, 22, and 25 of the IEEE 33-bus system. Each campus hosts an H100 GPU cluster together with a behind-the-meter (BtM) gas turbine, battery storage, and rooftop photovoltaics (PV); the configuration is summarized in Table~\ref{tab:dc_config}. Nodal electricity prices and carbon intensities are obtained by solving a per-period SOCP-ACOPF and the carbon emission flow model~\cite{kang2012cef} on the feeder, where the slack bus follows real-time CAISO LMPs and the network embeds distributed PV and gas units. \textcolor{black}{The carbon signals are computed using the baseline data center loads, while the effects of demand response on nodal carbon intensities are examined in Appendix C.}
The scheduling horizon spans four representative days at 15-min resolution. 

\begin{table}[!htbp]
\caption{Data Center Campus Configuration}
\label{tab:dc_config}
\centering
\footnotesize
\setlength{\tabcolsep}{7pt}
\begin{tabular}{@{}lccc@{}}
\toprule
 & \textbf{DC-A} & \textbf{DC-B} & \textbf{DC-C} \\
 & (Bus 18) & (Bus 22) & (Bus 25) \\
\midrule
GPUs         & 12,288 & 12,288 & 16,384 \\
PUE                        & 1.10 & 1.09 & 1.12 \\
\midrule
Turbine $P_{\max}/P_{\min}$ (kW) & 2000/500 & 2000/500 & 2000/500 \\
Turbine $\eta$ / fuel (\$/kWh)   & 0.30/0.025 & 0.30/0.025 & 0.30/0.025 \\
Battery $P$/$E$ (kW/kWh)         & 500/2500 & 500/2500 & 500/2500 \\
PV capacity (kW)                 & 600 & 800 & 400 \\
\bottomrule
\end{tabular}
\end{table}

\textbf{Scenarios.} As defined in Table~\ref{tab:scenarios}, six scenarios are designed to isolate the marginal contribution of spatial allocation (Sp), model switching (Sw), and quantization (Q), culminating in the full framework S6. 
In scenarios without quantization flexibility, each request tier follows a fixed default precision policy. 

\begin{table}[!htbp]
\caption{Scenario Definitions}
\label{tab:scenarios}
\centering
\footnotesize
\setlength{\tabcolsep}{6pt}
\begin{tabular}{@{}lcccccc@{}}
\toprule
 & \textbf{S1} & \textbf{S2} & \textbf{S3} & \textbf{S4} & \textbf{S5} & \textbf{S6} \\
\midrule
Spatial allocation & & \checkmark & & \checkmark & & \checkmark \\
Model switching    & & & \checkmark & \checkmark & \checkmark & \checkmark \\
Dynamic quantization       & & & & & \checkmark & \checkmark \\
DR participation   & & \checkmark & \checkmark & \checkmark & \checkmark & \checkmark \\
\midrule
Label & Base & Sp & Sw & Sw+Sp & Sw+Q & All \\
\bottomrule
\end{tabular}
\end{table}

\textbf{Model instances.} The serving library comprises three model families, Llama-3.1-405B, Qwen-2.5-72B, and DeepSeek-V3-671B, each available at four precision levels, FP16, INT8, GPTQ, and OmniQuant. Their key power-system parameters are summarized in Table~\ref{tab:instance_params} and are derived through the mapping of Section~\ref{sec:quantization_power_model}. Model architecture metadata are taken from the corresponding model reports~\cite{meta2024llama3,qwen2024techreport,deepseek2024v3}, while the hardware configuration follows NVIDIA specifications~\cite{nvidia2023h100,nvidia2022dgxh100}. 
The throughput values are calibrated from public LLM inference benchmarks and serving measurements, while the token-energy parameters are computed by the proposed quantization-to-power model and validated against reported power measurements~\cite{llminferencebench2024,niu2026tokenpowerbench,wattcounts2026,newkirk2025h100power,nvidia2023tensorrtllm}.
\textcolor{black}{The capacity redundancy factor $\rho$ is 0.9, and the calibration of its value is provided in Appendix \ref{app:capacity_headroom}.}
\textcolor{black}{The loading time $\tau_k^\ell$ accounts for checkpoint transfer and runtime initialization, and depends on model size, storage locality, and effective transfer bandwidth~\cite{fu2024serverlessllm}. The unloading time $\tau_k^u$ represents the delay required to release the outgoing instance's GPU resources.}
QoS degradation factors $\delta$ are calibrated from the quantization research results~\cite{dettmers2022llmint8,frantar2023gptq,shao2024omniquant,shi2025qmeter}. 

\textcolor{black}{
The per-GPU resident idle, loading, and unloading powers are set to 75, 150, and 50~W, respectively \cite{kocher2025energybenchmarks,niu2025energyengines}. For each kind of model, per-instance idle power is scaled by the number of occupied GPUs. Together with the loading and unloading times, these powers determine the switching energy. Model switching costs are captured through energy-related operating costs and the opportunity cost of temporarily unavailable serving capacity, represented by the service-capacity constraints.}

\textcolor{black}{Demand is grouped into three service tiers. 
Gold retains FP16 for strict-quality services, such as clinical decision support and high-stakes financial analysis~\cite{who2024lmm,finra2024genai}. 
Silver allows small quality losses, such as routine customer support and factual summarization. 
Bronze allows broader, bounded quality variation, such as low-risk dialogue, simple FAQs, and coarse text classification when availability takes priority.
The configuration-level degradation limits \(\delta_j^{\max}\) are set to 0, 0.06, and 0.08 for Gold, Silver, and Bronze.}
The demand shares of gold, silver, and bronze tiers are set to 48\%, 18\%, and 34\%~\cite{wang2025burstgpt}. Their QoS penalty coefficients are set to \$50, \$10, and \$3 per million tokens, respectively, following the relative prices of real-world GPT services.

\begin{table}[!htbp]
\caption{Core Model--Quantization Instance Settings. GPU/GB denotes GPUs per instance and total GPU memory.
$\mu_k$ is online-serving throughput, $e_k$ is per-token inference energy,
$\delta_k$ is QoS degradation relative to FP16,
\textcolor{black}{and $\tau_k^\ell$ and $\tau_k^u$ are model loading and unloading times, respectively.}}
\label{tab:instance_params}
\centering
\footnotesize
\setlength{\tabcolsep}{5pt}
\renewcommand{\arraystretch}{1.12}
\begin{tabular}{@{}llrrrrrr@{}}
\toprule
\textbf{Model} & \textbf{Quant.} &
\begin{tabular}[c]{@{}c@{}}\textbf{GPU}\\[-0.2ex]\textbf{/GB}\end{tabular} &
\begin{tabular}[c]{@{}c@{}}$\boldsymbol{\mu_k}$\\[-0.2ex]{\scriptsize tok/s}\end{tabular} &
\begin{tabular}[c]{@{}c@{}}$\boldsymbol{e_k}$\\[-0.2ex]{\scriptsize J/tok}\end{tabular} &
$\boldsymbol{\delta_k}$ &
\begin{tabular}[c]{@{}c@{}}\textcolor{black}{$\boldsymbol{\tau_k^\ell}$}\\[-0.2ex]{\scriptsize\textcolor{black}{s}}\end{tabular} &
\begin{tabular}[c]{@{}c@{}}\textcolor{black}{$\boldsymbol{\tau_k^u}$}\\[-0.2ex]{\scriptsize\textcolor{black}{s}}\end{tabular} \\
\midrule
\multirow{4}{*}{\begin{tabular}[c]{@{}l@{}}Llama-3.1\\405B\end{tabular}}
& FP16      & 12 / 933  &  650 & 8.00 & 0.00 & \textcolor{black}{750} & \textcolor{black}{90} \\
& INT8      &  6 / 467  &  750 & 4.50 & 0.01 & \textcolor{black}{500} & \textcolor{black}{75} \\
& GPTQ      &  4 / 241  &  700 & 3.50 & 0.03 & \textcolor{black}{360} & \textcolor{black}{54} \\
& OmniQuant &  3 / 183  &  600 & 3.00 & 0.07 & \textcolor{black}{300} & \textcolor{black}{45} \\
\midrule
\multirow{4}{*}{\begin{tabular}[c]{@{}l@{}}Qwen-2.5\\72B\end{tabular}}
& FP16      &  3 / 166  & 1000 & 1.40 & 0.00 & \textcolor{black}{120} & \textcolor{black}{18} \\
& INT8      &  2 / 84   & 1200 & 0.95 & 0.01 & \textcolor{black}{90}  & \textcolor{black}{14} \\
& GPTQ      &  1 / 44   &  800 & 0.75 & 0.03 & \textcolor{black}{70}  & \textcolor{black}{11} \\
& OmniQuant &  1 / 33   &  850 & 0.70 & 0.08 & \textcolor{black}{60}  & \textcolor{black}{10} \\
\midrule
\multirow{4}{*}{\begin{tabular}[c]{@{}l@{}}DeepSeek-V3\\671B\end{tabular}}
& FP16      & 20 / 1547 & 1200 & 8.00 & 0.00 & \textcolor{black}{900} & \textcolor{black}{90} \\
& INT8      & 10 / 776  & 1400 & 4.50 & 0.01 & \textcolor{black}{600} & \textcolor{black}{90} \\
& GPTQ      &  6 / 402  & 1300 & 3.50 & 0.03 & \textcolor{black}{480} & \textcolor{black}{72} \\
& OmniQuant &  4 / 306  & 1100 & 3.00 & 0.05 & \textcolor{black}{420} & \textcolor{black}{63} \\
\bottomrule
\end{tabular}
\end{table}

\textbf{Workload and prices.} The aggregate request profile is taken from the real-world BurstGPT trace~\cite{wang2025burstgpt}. Electricity prices use CAISO real-time LMPs~\cite{caiso2024oasis}. 
The carbon price is \$30/tCO$_2$, in line with California carbon allowance prices~\cite{carb2026auction}. 
\textcolor{black}{The DR program operates over a daily 19:00--21:00 window, with a capacity reward of \$2000/MWh, consistent with Southern California Edison's practice~\cite{sce_elrp}. 
For each scenario S2--S6, the DR baseline is obtained from the corresponding optimized no-DR trajectory. 
Each paired comparison uses identical operating capabilities, exogenous inputs, and initial conditions.}

\textcolor{black}{
\textbf{Implementation and solver settings.}
Instances resident at the beginning of the scheduling horizon are assumed to be loaded and ready to serve. Their pre-horizon startup energy and orchestration costs are excluded.
S1 and S2 retain their prescribed static instance configurations. For S3--S6, the initial resident configuration is selected by the corresponding no-DR optimization and then fixed in the paired DR run. Each battery starts at 50\% of its rated energy capacity.
The optimization model is implemented in Python and solved using Gurobi 13.0.2 on an Intel Xeon Platinum 8468V system. The target relative MIP gap is 0.1\%.} 

\begin{figure}[!htbp]
\centering
\includegraphics[width=0.9\columnwidth]{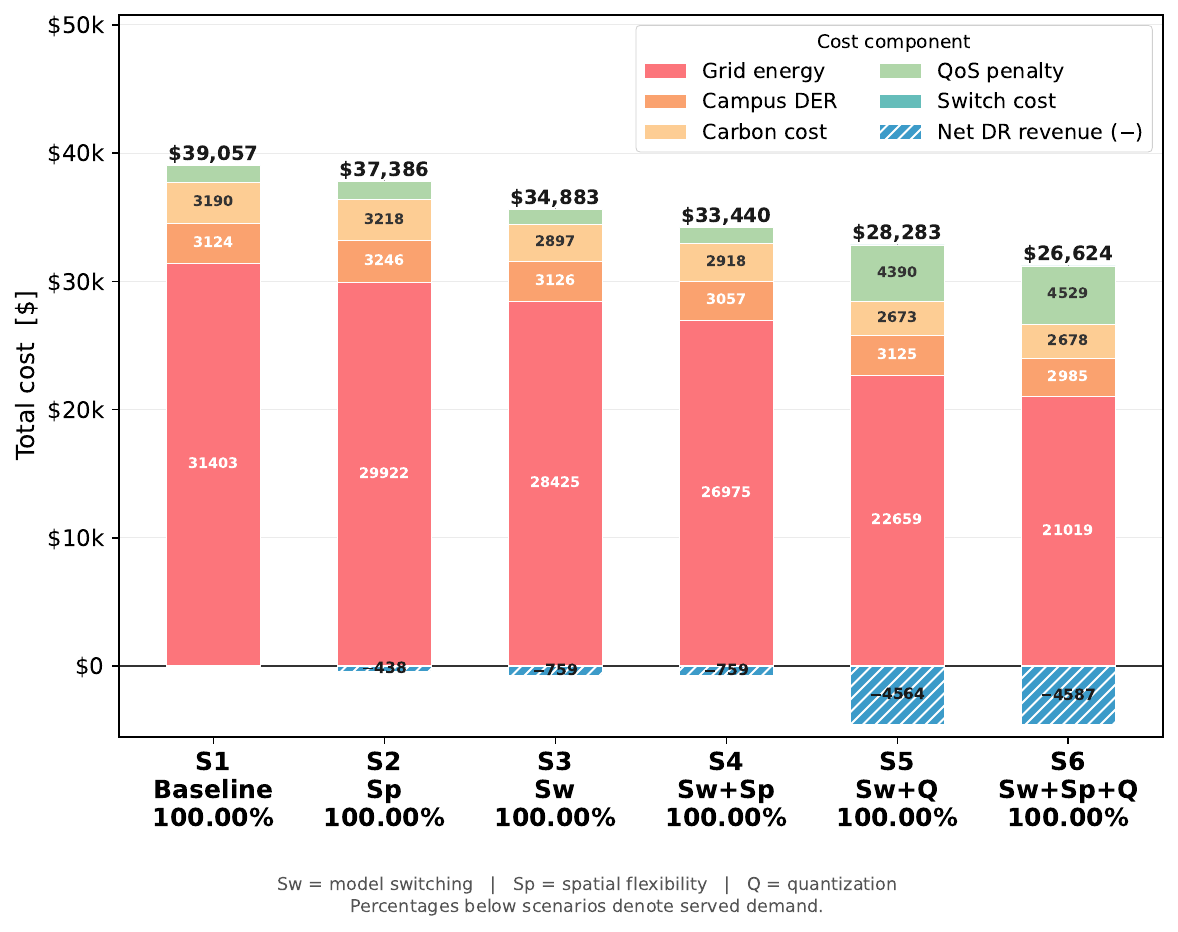}
\caption{Daily cost composition by flexibility scenario (S1--S6). Sw, Sp, and Q denote model switching, spatial flexibility, and quantization, respectively.}
\label{fig:cost_by_scenario}
\end{figure}

\subsection{Overall Performance and Campus Energy Operation}
\label{sec:result_overall}

\begin{figure}[!b]
\centering
\subfloat[S1 (baseline)]{%
    \includegraphics[width=\columnwidth]{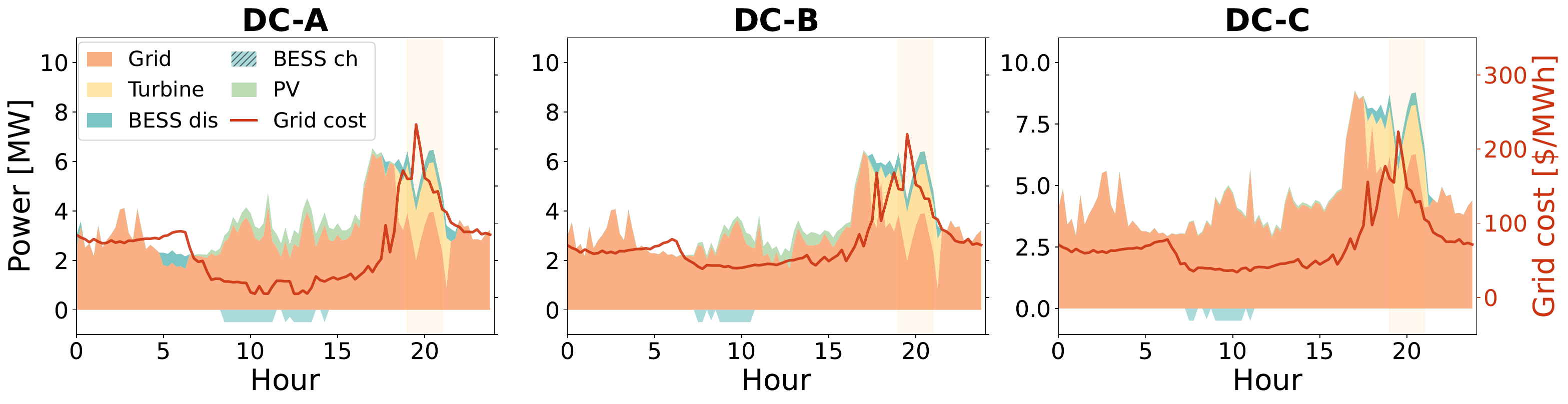}%
    \label{fig:power_composition_s1}}
\\
\subfloat[S4 (switching + spatial)]{%
    \includegraphics[width=\columnwidth]{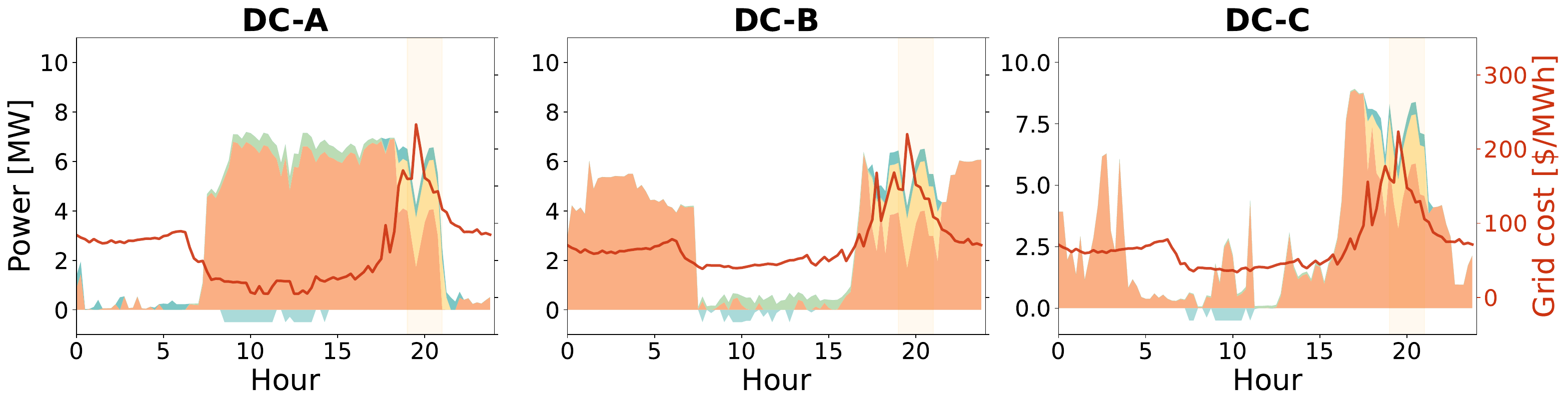}%
    \label{fig:power_composition_s4}}
\\
\subfloat[S6 (full framework)]{%
    \includegraphics[width=\columnwidth]{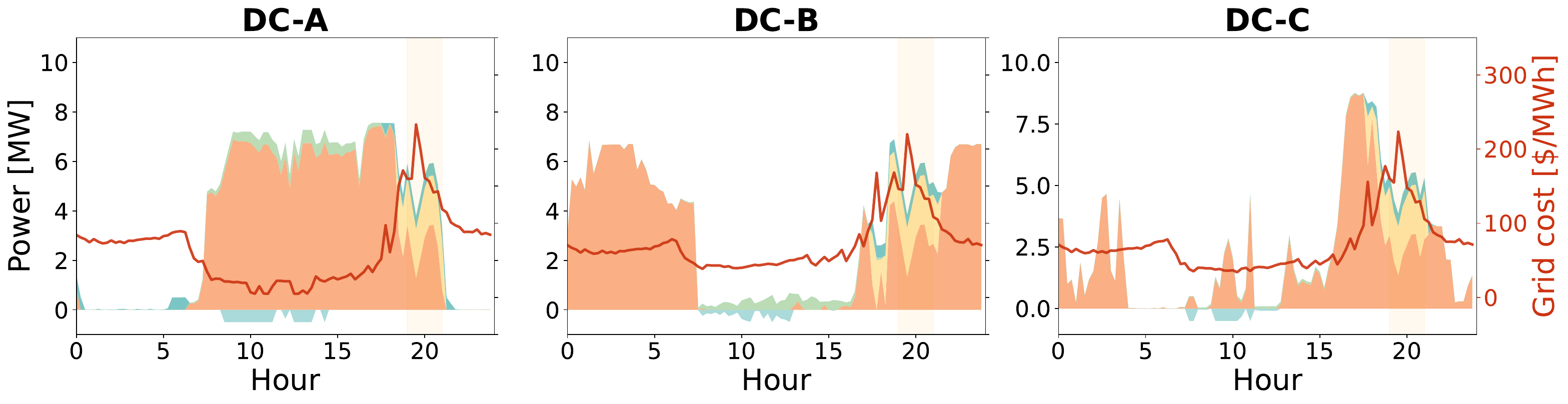}%
    \label{fig:power_composition_s6}}
\caption{Per-campus power supply composition over a representative day under S1, S4 and S6. The DR window is shaded.}
\label{fig:power_composition}
\end{figure}

\textcolor{black}{Fig.~\ref{fig:cost_by_scenario} decomposes the average daily operating cost of the three campuses across the six scenarios, calculated by dividing the total cost over the four-day scheduling horizon by four.
For clarity, Figs.~\ref{fig:power_composition}--\ref{fig:precision_throughput} show the operating trajectories for Day~1 of the same four-day solution; the corresponding results for Days~2--4 are provided in Appendix~\ref{app:multiday_results}.}

Three observations can be made from Fig.~\ref{fig:cost_by_scenario}.
First, the proposed full framework in S6 reduces the total operating cost by \textcolor{black}{31.8\%} compared with the no-flexibility baseline S1. This reduction is mainly driven by lower grid electricity purchase cost, lower carbon-related cost, and increased DR compensation, demonstrating the overall effectiveness of the proposed quantization-enabled coordinated operation framework.
Second, all flexibility mechanisms contribute to cost reduction, but their marginal impacts differ. Among them, quantization provides the most significant benefit. Compared with S4, S6 further reduces the total cost by \textcolor{black}{20.4\%}, showing that precision-tunable flexibility is a promising demand-side resource. 
Third, stacking different flexibility mechanisms leads to a clear further cost reduction. For example, S4 achieves lower cost than both S2 and S3, and S6 further outperforms both S4 and S5. This result shows that the proposed quantization-enabled flexibility is highly compatible with conventional spatial workload shifting, and can further amplify the demand-response capability of data centers.

Fig.~\ref{fig:power_composition} shows the per-campus power supply composition over a representative day. 
The battery charges during the daytime low-price and PV-rich period and discharges during the evening peak, while the turbine operates during high-price evening periods, including the DR window.
Compared with S1, S6 substantially lowers the overall power level and reduces dependence on grid import, especially during the evening DR window.
These results show that IT-side flexibility improves the utilization of local energy assets.
Model switching and quantization reduce the underlying inference power demand, allowing the on-site turbine and BESS to cover a larger share of the residual demand during high-price DR periods. Thus, computing-side flexibility amplifies the value of campus energy resources in grid-interactive operation.

\subsection{Quantization-Enabled Inference Flexibility}
\label{sec:result_quan}

\begin{figure}[!b]
\centering
\subfloat[S4 (switching + spatial, no quantization)]{%
    \includegraphics[width=0.42\textwidth]{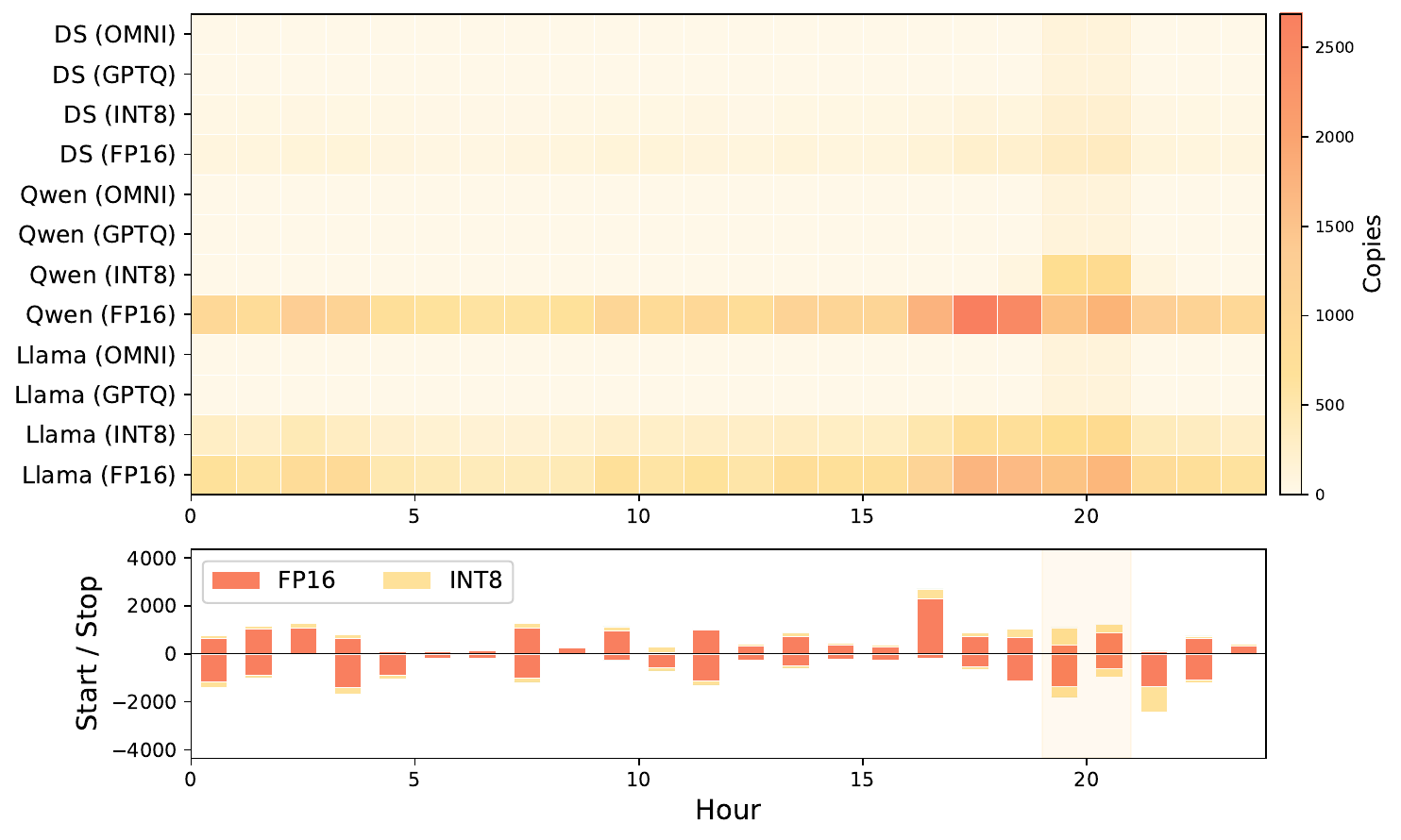}%
    \label{fig:instance_switching_s4}}
\\
\subfloat[S6 (full framework, with quantization)]{%
    \includegraphics[width=0.42\textwidth]{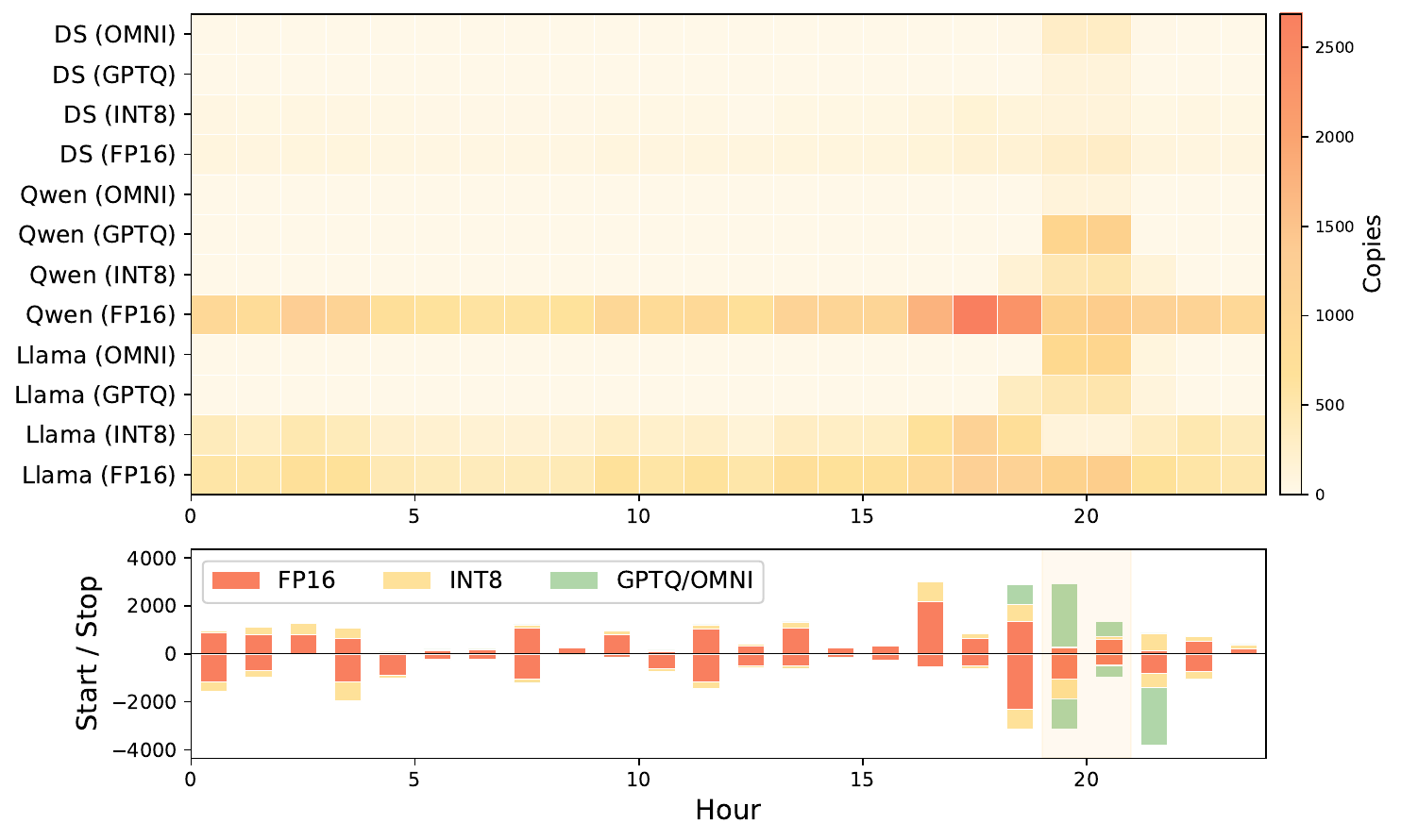}%
    \label{fig:instance_switching_s6}}
\caption{Model-instance switching under S4 and S6. The top of each panel represents loaded copies of each model summed over the three campuses; the bottom represents startup and shutdown actions.}
\label{fig:instance_switching}
\end{figure}

Fig.~\ref{fig:instance_switching} illustrates how quantization changes the model-instance loading strategy by comparing S4 and S6, with additional quantization flexibility enabled in S6. 
In S4, the loaded instances are limited to FP16 and INT8, while GPTQ and OmniQuant remain inactive throughout the day. 
In contrast, S6 selectively activates GPTQ and OmniQuant instances, but their startups and shutdowns are mainly concentrated around the high-price evening period and the DR window.
This pattern indicates that the proposed method does not use GPTQ and OmniQuant as default serving modes. Instead, it weighs the QoS loss against the electricity price and DR value, and activates low-precision instances only when their flexibility is economically justified. 
Therefore, quantization serves as a complementary control lever, where high-precision instances maintain normal service quality, while low-precision instances are introduced when the campuses need to reduce LLM inference power and respond to grid signals.

\begin{figure}[!htbp]
\captionsetup[subfloat]{aboveskip=-2pt}
\centering
\subfloat[{\small S1 (baseline, no quantization flexibility)}]{%
    \includegraphics[width=0.42\textwidth]{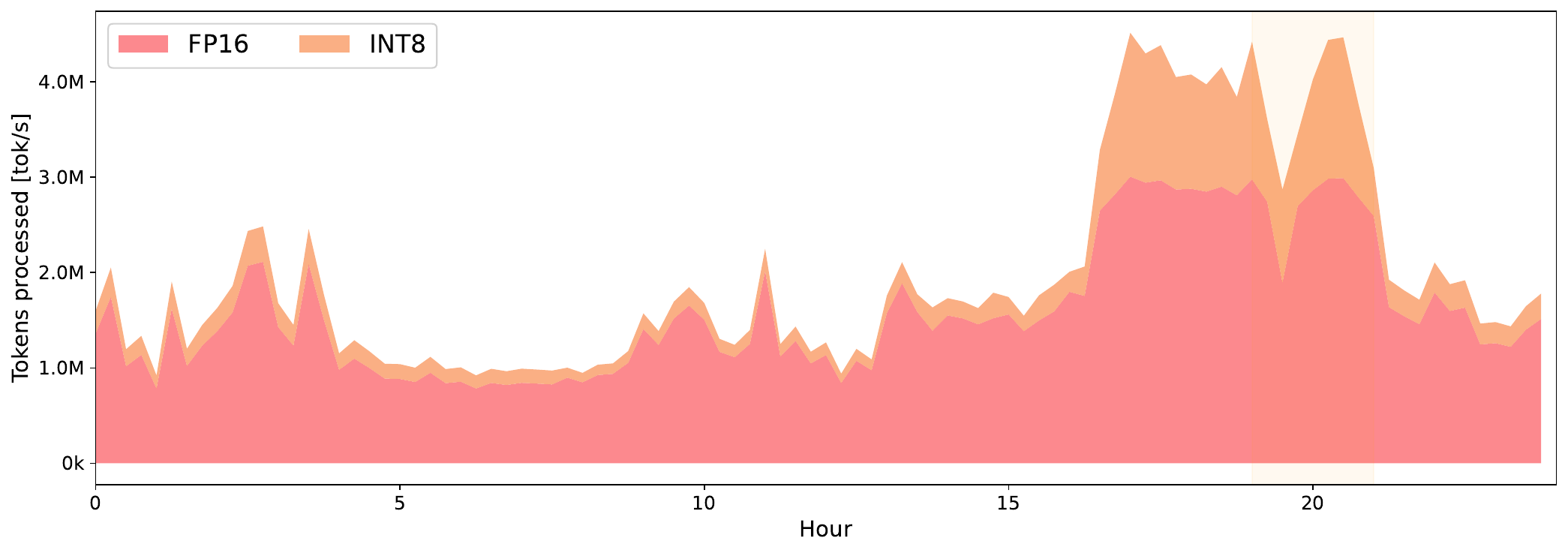}%
    \label{fig:precision_throughput_s1}}
\\
\subfloat[{\small S6 (full framework, with quantization flexibility)}]{%
    \includegraphics[width=0.42\textwidth]{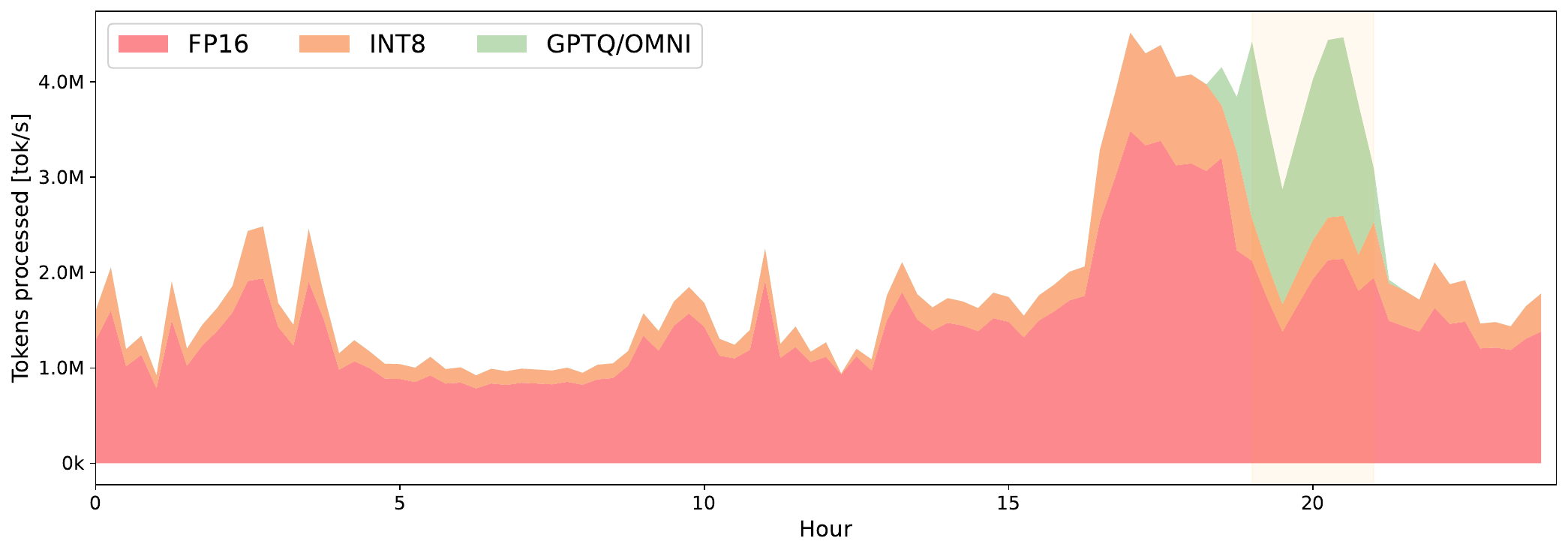}%
    \label{fig:precision_throughput_s6}}
\caption{Token throughput by serving precision over a representative day under (a) S1 and (b) S6. The INT8 component in S1 reflects the static default precision of bronze-tier requests.}
\label{fig:precision_throughput}
\end{figure}

Fig.~\ref{fig:precision_throughput} shows how the served token throughput is distributed across precision levels. 
The total served throughput remains unchanged across scenarios, indicating that quantization does not reduce or defer inference demand, but changes the precision at which requests are served. 
In S1, INT8 is mainly used for bronze-tier requests, while lower-precision serving remains limited under normal operating conditions. In S6, GPTQ and OmniQuant are activated primarily during the high-cost evening period.
This result highlights the role of quantization as a routing-based flexibility mechanism. 
Under normal conditions, only the bronze traffic is served at reduced precision to limit QoS degradation. When electricity cost and DR value increase, the proposed model extends quantized serving to a larger share of eligible traffic, including silver-tier requests, to reduce LLM inference power without curtailing demand.

\subsection{Sensitivity Analysis}
\label{sec:result_sensitivity}

\begin{figure}[!htbp]
\centering
\includegraphics[width=0.85\columnwidth]{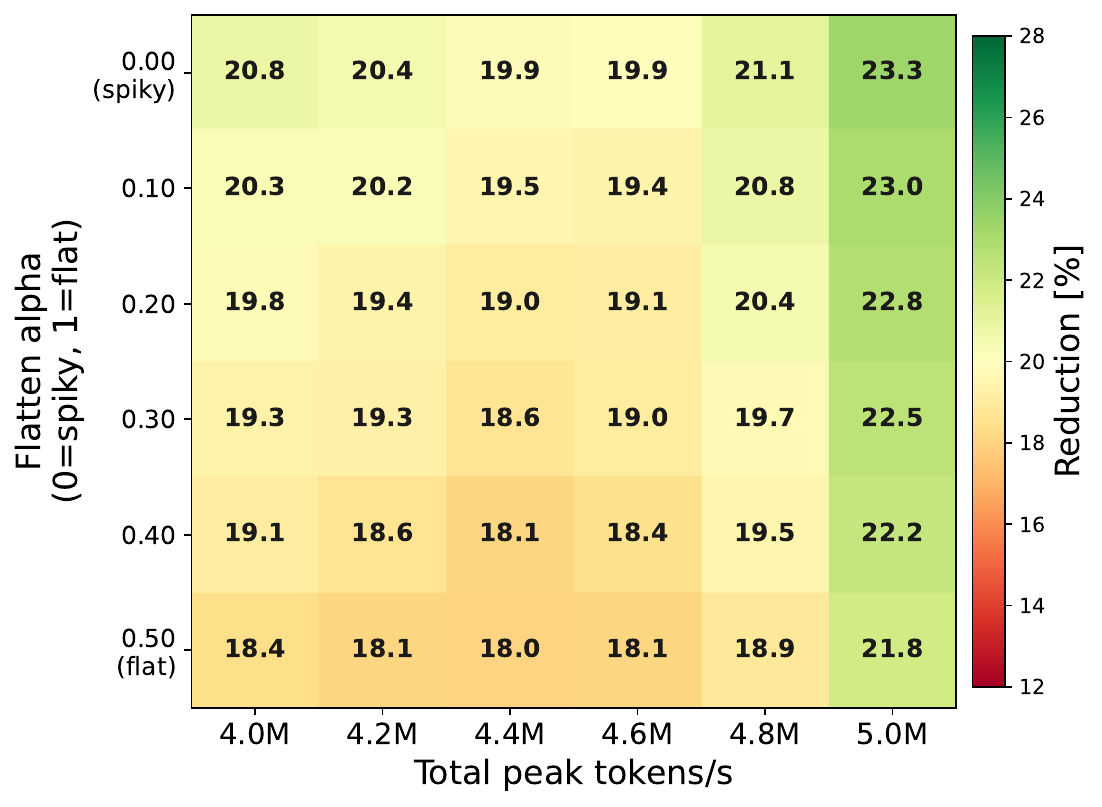}
\caption{Sensitivity of the quantization benefit to workload peak rate and flattening factor \(\alpha\). Here, \(\alpha=0\) denotes the original spiky profile and \(\alpha=0.5\) denotes a flatter profile. The full-service requirement is relaxed to allow unmet demand when serving capacity is insufficient, with unserved-token penalties included in the objective.}
\label{fig:workload_sensitivity}
\end{figure}

To assess the robustness of the quantization benefit, Fig.~\ref{fig:workload_sensitivity} varies the workload along two dimensions: the peak request rate and a flattening factor \(\alpha\), which interpolates the load profile from spiky \((\alpha=0)\) to flat \((\alpha=0.5)\). 
The figure quantifies the additional value of quantization relative to the conventional-flexibility scenario S4.
Fig.~\ref{fig:workload_sensitivity} reports the relative cost reduction of S6 over S4, which is the extra saving obtained from quantization. 
The reduction remains substantial across all tested workloads, ranging from \textcolor{black}{18.0\% to 23.3\%}, which confirms the robust value of quantization. 
\textcolor{black}{At the highest peak rate of 5.0 million tokens/s, S6 reduces total operating cost by 21.8\%–23.3\% relative to S4. 
This advantage partly arises because quantization alleviates serving-capacity pressure, allowing S6 to serve more requests and incur lower unserved-token penalties than S4. 
These results demonstrate that quantization-enabled flexibility can both reduce operating costs and maintain service rate under high-load conditions.}

\begin{figure}[!htbp]
\centering
\includegraphics[width=0.8\columnwidth]{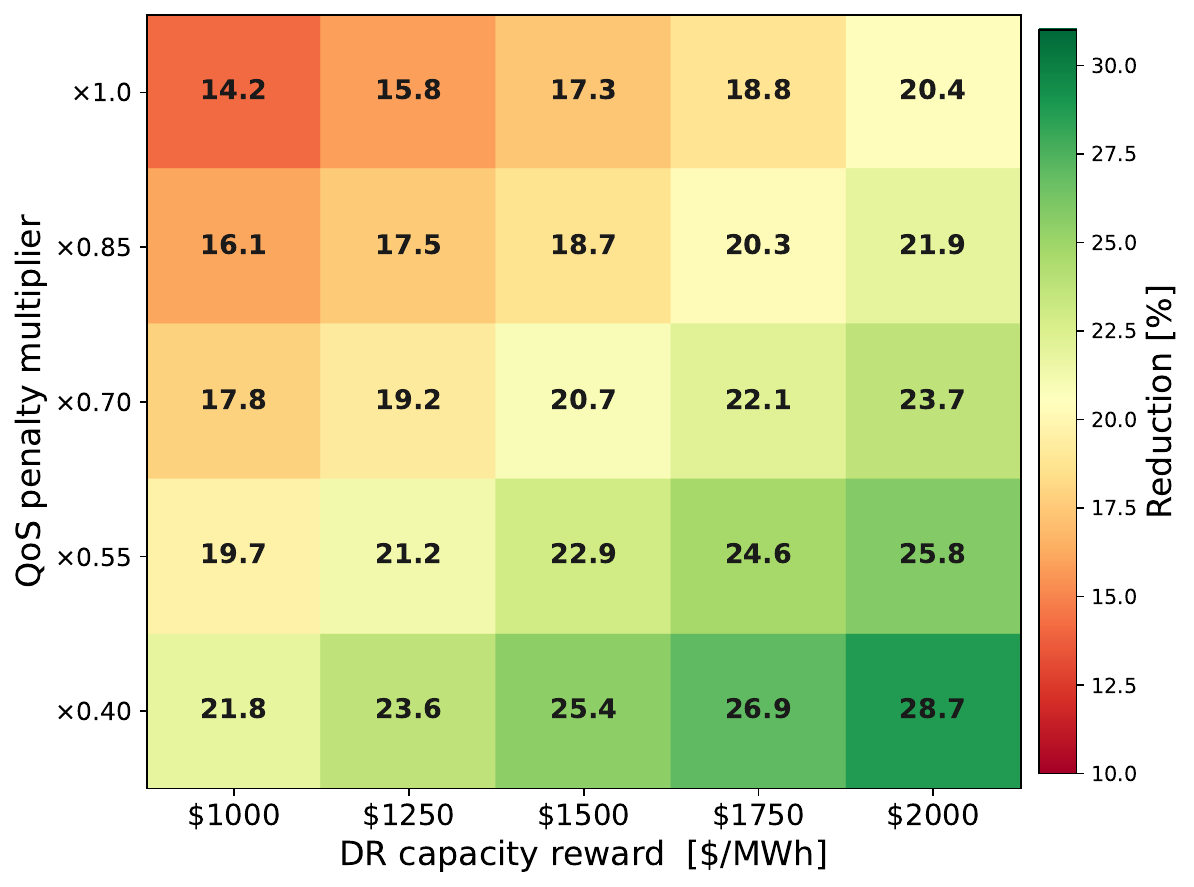}
\caption{Cost reduction of S6 over S4 across the price plane of DR capacity reward and QoS penalty multiplier. Here, \(\times 1.0\) denotes the baseline QoS penalty setting calibrated from GPT service prices.}
\label{fig:price_plane_sensitivity}
\end{figure}

\begin{figure*}[!htbp]
\color{black}
\centering
\captionsetup{font=small}
\subfloat[\textcolor{black}{Service-tier composition}]{%
\includegraphics[width=0.33\textwidth]{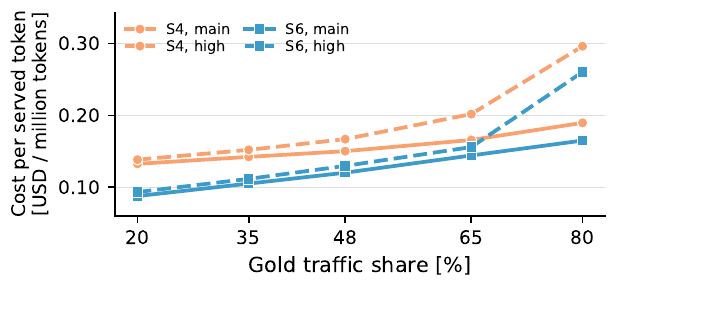}}\hfill
\subfloat[\textcolor{black}{Degradation tolerance}]{%
\includegraphics[width=0.33\textwidth]{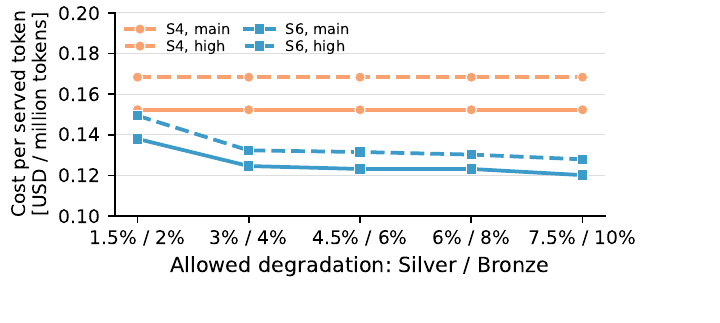}}\hfill
\subfloat[\textcolor{black}{Quality-penalty coefficient}]{%
\includegraphics[width=0.33\textwidth]{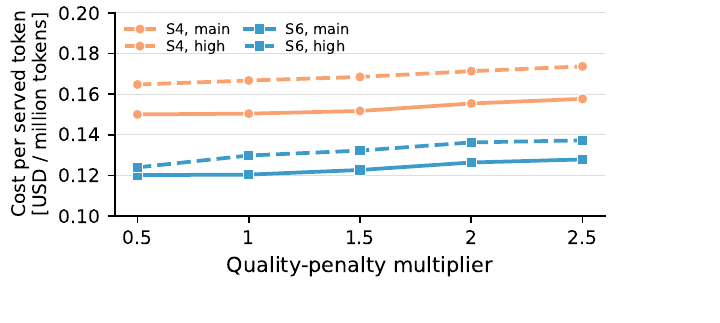}}
\caption{\textcolor{black}{Sensitivity of cost per million served tokens to (a) service-tier composition, (b) degradation tolerance, and (c) the quality-penalty coefficient. The high-load scenario uses 1.1$\times$ the main-scenario workload. The full-service requirement is relaxed to allow unmet demand when serving capacity is insufficient, with unserved-token penalties included in the objective.}}
\label{fig:service_sensitivity}
\end{figure*}

Fig.~\ref{fig:price_plane_sensitivity} maps the cost reduction of S6 over S4 on the price plane defined by the DR reward and the QoS penalty. Here, the QoS penalty represents the cost of degrading service quality through quantization.
Across all tested price settings, quantization provides substantial additional flexibility. Even under the most unfavorable tested condition, with the highest QoS penalty and the lowest DR reward, S6 still reduces the total cost by \textcolor{black}{14.2\%} compared with S4. 
As the QoS penalty decreases, the cost reduction becomes more pronounced, indicating that the economic value of quantization is sensitive to service-quality degradation cost. 
\textcolor{black}{Increasing the DR reward also increases the cost reduction and the cost advantage persists even at the lowest DR reward tested.}
Therefore, as long as moderate QoS degradation is acceptable for part of the workload, quantization can provide a robust and cost-effective flexibility resource for LLM data centers.

\textcolor{black}{
Fig.~\ref{fig:service_sensitivity} compares the cost per served token of S4 and S6 under varying service-tier compositions, degradation limits, and quality-penalty coefficients. 
Three observations follow.
First, quantization helps alleviate serving-capacity pressure under high workloads. 
At a 65\% Gold share, S6 increases the token service rate from 99.28\% to 99.97\% by exploiting precision flexibility for eligible requests.
Second, substantial savings are achievable with modest degradation allowances. 
Beyond the strictest limits, admitting additional quantized configurations enables larger cost reductions across the remaining tested range.
Third, the cost advantage is robust to higher quality penalties. 
Even at 2.5 times the reference coefficients, S6 reduces the cost per served token by 18.91\% and 20.97\% under the main and higher workloads.
}

\begingroup
\color{black}
\subsection{\textcolor{black}{Robustness to Load Forecasting Errors}}
\label{sec:forecast_robustness}

To assess robustness to forecasting errors, we first optimized S6 using forecast loads, with 10\% capacity redundancy ($\rho=0.9$) or without redundancy ($\rho=1.0$). 
We then independently perturbed each forecast value using zero-mean Gaussian relative errors, with standard deviations $\sigma$ of 0--10\%, clipped at $\pm3\sigma$. 
Each error level includes ten four-day load curves, yielding 40 daily observations. 
We evaluated the perturbed loads with instance schedules fixed, while allowing request routing and continuous energy dispatch to adjust.
Both settings use the full physical service capacity during evaluation.

Fig.~\ref{fig:forecast_robustness} compares the token service-rate distributions with and without capacity redundancy across forecast-error levels. 
With redundancy, the mean service rate remains above 99\% throughout, reaching 99.34\% at $\sigma=10\%$, compared with 96.61\% without redundancy. 
Reserving 10\% of service capacity increases scheduling cost per token by only 3.33\% relative to the setting without redundancy. 
Thus, capacity redundancy mitigates service losses from forecasting errors at a modest additional cost, effectively improving the robustness of the proposed method.

\begin{figure}[!tbp]
\color{black}
\centering
\captionsetup{font=small}
\includegraphics[width=\columnwidth]{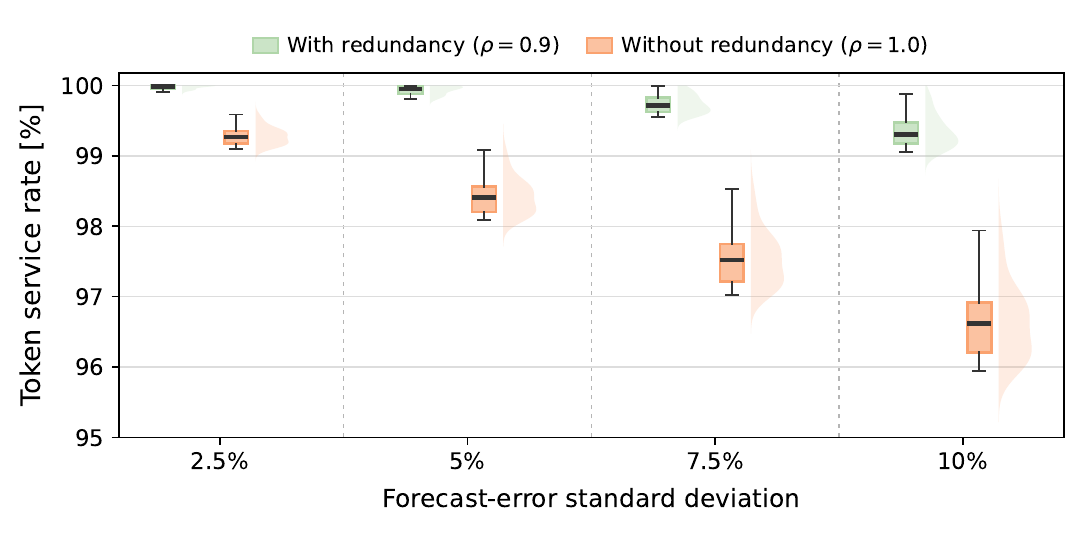}
\caption{\textcolor{black}{Token service rates under forecasting errors in S6.}}
\label{fig:forecast_robustness}
\end{figure}

\endgroup

\section{Conclusion}
\label{sec:conclusion}

This paper introduced LLM quantization as an IT-side flexibility lever for grid-responsive LLM data-center energy management, resolving inference flexibility down to the model-instance level.
The proposed framework consists of a quantization-to-power mapping, a two-stage DR model, and a multi-campus co-optimization method.
Case studies based on real-world data lead to three conclusions.
First, quantization provides a distinct and additive flexibility lever.
Enabling quantization further reduces the total cost by \textcolor{black}{20.4\%} beyond conventional model switching and spatial flexibility, and its benefit remains positive under all tested workload and price conditions.
Second, this benefit is achieved without reducing the served token volume. The framework routes requests to lower precision only when electricity cost and DR value justify the bounded quality loss.
Third, IT-side and energy-side flexibility reinforce each other. By reducing the underlying inference load, quantization enables limited on-site assets to cover a larger share of campus demand during DR events.
Quantization therefore remains a useful flexibility resource for the large and sustained inference loads of future LLM data centers.

\FloatBarrier 
\appendices
\setcounter{equation}{0}
\renewcommand{\theequation}{A\arabic{equation}}
\renewcommand{\theHequation}{appendix.\arabic{equation}}
\begingroup
\setcounter{figure}{0}
\setcounter{table}{0}
\renewcommand{\thefigure}{A\arabic{figure}}
\renewcommand{\thetable}{A\arabic{table}}
\renewcommand{\theHfigure}{appendix.\arabic{figure}}
\renewcommand{\theHtable}{appendix.\arabic{table}}
\color{black}
\captionsetup[figure]{font=small}
\setlength{\intextsep}{6pt plus 1pt minus 1pt}
\setlength{\textfloatsep}{8pt plus 2pt minus 2pt}
\section{Supplementary Evaluation of Quantization Flexibility}
\label{app:quantization_flexibility}

\textit{Experimental setup:}
This experiment assesses whether quantization provides additional energy savings alongside frequency control \cite{wang2025loadflexibility} and power capping \cite{patel2024asplos}. We deployed DeepSeek-R1-Distill-Llama-70B using vLLM on NVIDIA A100 GPUs, comparing one BF16 instance with one 8-bit instance. Each configuration processed the same benchmark prompts.
We evaluated default serving, frequency control at 1095~MHz, and power capping at 225~W per GPU. Within each setting, both precisions used identical control parameters.
Energy was measured over the complete workload execution and summed across the GPUs assigned to each instance, including their idle power during execution.

\begin{table}[!htbp]
\color{black}
\centering
\caption{GPU energy consumption and reduction rate under quantization and other control methods.}
\label{tab:appendix_quantization_energy}
\small
\setlength{\tabcolsep}{3pt}
\renewcommand{\arraystretch}{1.25}
\begin{tabular}{@{}
    >{\raggedright\arraybackslash}m{0.31\linewidth}
    >{\centering\arraybackslash}m{0.19\linewidth}
    >{\centering\arraybackslash}m{0.19\linewidth}
    >{\centering\arraybackslash}m{0.19\linewidth}
@{}}
\hline
Control setting
& BF16\newline(kJ)
& W8A8\newline(kJ)
& Reduction\newline(\%) \\
\hline
Default serving   & 1882.6 & 1085.6 & \textbf{42.3} \\
Frequency control & 1339.5 &  726.2 & \textbf{45.8} \\
Power capping     & 1580.9 &  805.2 & \textbf{49.1} \\
\hline
\end{tabular}
\end{table}

\textit{Results:}
As shown in Table~\ref{tab:appendix_quantization_energy}, 8-bit quantization reduces GPU energy consumption by 42.3\% under default serving and retains reductions of 45.8\% and 49.1\% with frequency control and power capping, respectively.
These observations support precision selection and its associated GPU allocation as an additional configuration choice that complements existing power controls.

\section{Validation of Interval-Level Throughput and Energy Approximation}
\label{app:equivalent_energy_validation}

\textit{Experimental setup:}
We evaluated the stability of interval-level average throughput and equivalent active energy on a local server equipped with one NVIDIA A100 80~GB GPU. 
The experiments used DeepSeek-R1-Distill-Llama-8B in BF16 precision using vLLM. 
The prompt set combined requests from AlpacaEval, Arena-Hard, WildBench, and a ShareGPT-derived prompt pool \cite{li2023alpacaeval,li2025arenahard,lin2024wildbench}. 
These sources cover instruction-following tasks, challenging open-ended questions, and real-world conversations with diverse prompt and response lengths. 

\textit{Measurement procedure:}
Three 15-min runs were conducted with the model and serving configuration held unchanged. 
In each run, prompts were randomly sampled from the benchmark sources. 
Each batch contained 128 requests and was submitted immediately after the preceding batch completed.
Average output throughput was measured as $\widehat{\mu}_t=N_t^{\mathrm{out}}/T_t$, where $T_t$ is the accumulated batch-generation time in seconds, including both
prefill and decode.
GPU energy was recorded using the NVML cumulative energy counter over complete generation runs, thereby covering both prefill and decode. 
The loaded-idle baseline was measured separately and subtracted. For each window \(t\), we calculated
\begin{equation}
    \rho_t =
    \frac{N_{t}^{\mathrm{in}}}{N_{t}^{\mathrm{out}}},
    \qquad
    e_{k,t}^{\mathrm{eq}} =
    \frac{E_{k,t}^{\mathrm{meas}}
    -p_k^{\mathrm{idle}}T_t}
    {N_{t}^{\mathrm{out}}},
    \label{eq:appendix_equivalent_energy}
\end{equation}
where \(\rho_t\) is the input-to-output token ratio, \(E_{k,t}^{\mathrm{meas}}\) is the measured GPU energy, and \(e_{k,t}^{\mathrm{eq}}\) is the loaded-idle-subtracted active energy
normalized by the number of output tokens.

\begin{table}[!htbp]
\color{black}
    \caption{Interval-level average output throughput, input-to-output token ratios, and equivalent active-energy coefficients.}
    \label{tab:equivalent_energy_validation}
    \centering
    \footnotesize
    \setlength{\tabcolsep}{3.2pt}
    \begin{tabular}{lccccc}
        \toprule
        Metric & Window 1 & Window 2 & Window 3 & Mean & CV (\%) \\
        \midrule
        $\widehat{\mu}_t$ (token/s)
        & 1418.45 & 1421.43 & 1420.88 & 1420.25 & 0.11 \\
        $\rho_t$
        & 4.6716 & 4.6768 & 4.6334 & 4.6606 & 0.51 \\
        $e_{k,t}^{\mathrm{eq}}$ (J/token)
        & 0.145356 & 0.145157 & 0.145310 & 0.145274 & 0.07 \\
        \bottomrule
    \end{tabular}
\end{table}

\textit{Results for throughput stability:}
The three windows process 39,168 requests and generate 3,846,321 output tokens.
As summarized in Table~\ref{tab:equivalent_energy_validation}, the average output throughput has a between-window coefficient of variation of approximately 0.11\%.
Despite differences in the sampled requests and their lengths, window-level throughput remains stable. 
This provides empirical support for using a fixed reference throughput parameter for 15-min capacity scheduling.

\textit{Results for the equivalent-energy approximation:}
The input-to-output token ratio has a between-window coefficient of variation of 0.5089\%, and the equivalent active-energy coefficient exhibits a coefficient of variation of 0.0717\%.
These results support the interval-level energy approximation under the tested serving configuration. 
After aggregation over 15 minutes, the average prefill contribution can be incorporated into \(e_k\), while output tokens serve as the common workload-accounting unit for the dispatch model.

\begingroup
\color{black}
\hypersetup{linkcolor=black,citecolor=black,urlcolor=black}
\section{\textcolor{black}{Feedback of Data Center Demand Response on Carbon Signals}}
\label{app:carbon_feedback}

Data center demand response may affect grid dispatch and power flows, thereby changing nodal carbon intensities. 
To quantify the resulting effect on data center emissions, we first solved S1 and used its net grid imports in the ACOPF and carbon emission flow calculations to obtain the baseline nodal signals. 
S6 was then optimized once under these fixed signals, following the formulation adopted in this paper. 
The emissions calculated for this schedule using the input carbon intensity signals are termed \textit{estimated emissions}. 
We then substituted the optimized net grid imports into the grid model to recompute dispatch and carbon flows. 
The emissions recalculated for the same schedule are termed \textit{recalculated emissions}. 
Both quantities include grid-attributed emissions and direct campus emissions.

Table~\ref{tab:carbon_feedback} compares the two emission results. 
Estimated and recalculated emissions are 71.5091 and 71.4474~tCO$_2$/day, respectively, with an absolute relative difference of only 0.0864\%. 
This small discrepancy indicates that feedback has a limited effect on the aggregate emissions estimate in this case.

\begin{table}[!htbp]
\color{black}
\centering
\footnotesize
\caption{Comparison of Estimated and Recalculated Data Center Emissions for S6}
\label{tab:carbon_feedback}

\renewcommand{\arraystretch}{1.3}
\setlength{\tabcolsep}{8pt}
\begin{tabular}{lc}
\hline
Metric & S6 \\
\hline
Estimated emissions (tCO$_2$/day) & 71.5091 \\
Recalculated emissions (tCO$_2$/day) & 71.4474 \\
Absolute relative difference (\%) & 0.0864 \\
\hline
\end{tabular}

\smallskip
\begin{minipage}{0.95\linewidth}
\footnotesize
The absolute relative difference is calculated as
$|E_{\mathrm{DC}}^{\mathrm{est}}-
E_{\mathrm{DC}}^{\mathrm{rec}}|
/E_{\mathrm{DC}}^{\mathrm{rec}}\times100\%$,
where $E_{\mathrm{DC}}^{\mathrm{est}}$ and
$E_{\mathrm{DC}}^{\mathrm{rec}}$ denote estimated and
recalculated emissions.
\end{minipage}
\end{table}
\endgroup

\section{\textcolor{black}{Effect of Quantization on Output Length}}
\label{app:quantization_length_subset}

\textcolor{black}{
\textit{Experimental setup:}
We deployed DeepSeek-R1-Distill-Llama-8B on an NVIDIA A100 GPU using vLLM, comparing BF16, LLM.int8, GPTQ, and OmniQuant. Our analysis used 750 questions drawn from AlpacaEval~2~\cite{li2023alpacaeval}, Arena-Hard~v2~\cite{li2025arenahard}, and WildBench~v2~\cite{lin2024wildbench}.
Each configuration received the same questions and used identical decoding settings: temperature $0.6$, top-$p$ $0.95$, and a fixed random seed. We recorded output token counts and compared their means against BF16.}

\begin{table}[!htbp]
\color{black}
    \centering
    \footnotesize
    \begingroup
    \color{black}
    \setlength{\tabcolsep}{5pt}
    \caption{\textcolor{black}{Effect of Model Quantization on Output Length.}}
    \label{tab:quantization_length}
    \begin{tabular}{lcc}
        \toprule
        Configuration & Mean output (tokens) & Output change (\%)   \\
        \midrule
        BF16             & 2060.68 & ---      \\
        LLM.int8         & 2055.44 & $-0.25$  \\
        GPTQ             & 1950.22 & $-5.36$  \\
        OmniQuant        & 1989.88 & $-3.44$  \\
        \bottomrule
    \end{tabular}
    \endgroup
\end{table}

\textcolor{black}{
\textit{Results:}
As shown in Table~\ref{tab:quantization_length}, mean output lengths decrease by $0.25\%$--$5.36\%$ compared with the original precision. These results indicate that quantization affects output length, but the changes are slight. KV-cache memory consumption scales approximately linearly with the total sequence length. Therefore, the changes in output length have only a modest effect on GPU memory requirements in Eq.~\eqref{eq:gpu_memory_limit}. 
Based on these results, this paper assumes that output length remains approximately unchanged across quantization configurations at the scheduling level. This assumption avoids excessive resource reservation and also preserves a small margin for practical operation.}

\section{Request-Level Evaluation of Capacity Redundancy}
\label{app:capacity_headroom}

\textit{Experimental setup:}
We deployed two DeepSeek-R1-Distill-Llama-8B instances in BF16 using vLLM, with each replica running on one NVIDIA A100 80~GB GPU. Prompts were drawn from AlpacaEval~2, Arena-Hard~v2, and WildBench~v2~\cite{li2023alpacaeval,li2025arenahard,lin2024wildbench}. The request latency constraints are time to first token (TTFT) $\leq 2.5$~s and request-average time per output token (TPOT) $\leq 50$~ms. 

\textit{Measurement procedure:}
Firstly, we measured the maximum throughput $\mu_{ref}$ to be 475 output tokens/s based on the latency constraints.
Secondly, we selected three 15-min windows from BurstGPT~\cite{wang2025burstgpt}. For each window $w$, its second-resolution load $s_w(\tau)$ was scaled as
\begin{equation}
    \ell_w(\tau)
    = \rho\mu_{\mathrm{ref}}
      \frac{s_w(\tau)}{\bar{s}_w},
\end{equation}
where $\bar{s}_w$ is the original window-average load. Normalization by $\bar{s}_w$ removes the original workload scale while preserving relative fluctuations and their timing.
Multiplication by $\rho\mu_{\mathrm{ref}}$ then sets the target average load without compressing the 15-min window. 
Benchmark prompts were scheduled to approximate this request arrival profile. 
At $\rho=1$, the average offered load equals the reference capacity, whereas $\rho<1$ reserves a fraction $1-\rho$ as capacity redundancy.
For example, $\rho=0.90$ corresponds to an average load of 427.5 output tokens/s.
Comparing request-level latency across $\rho$ values therefore assesses the margin needed to accommodate the short-term fluctuations.

\textit{Results:}
All three windows pass at $\rho=0.85$ and $0.90$, whereas one fails at $\rho=0.95$.
These results support $\rho=0.90$, corresponding to 10\% redundancy.


\section{\textcolor{black}{S6 Operating Results for Days 2--4}}
\label{app:multiday_results}

\textcolor{black}{Figs.~\ref{fig:multiday_power_cost}--\ref{fig:multiday_token_rates} summarize the S6 trajectories for Days~2--4 from the same four-day solution. 
Their main operating patterns are consistent with Day~1, so these additional results are provided here to keep the main text concise.}

\begin{figure}[!htbp]
\color{black}
\centering
\captionsetup{font=footnotesize,skip=2pt}
\includegraphics[width=\columnwidth]{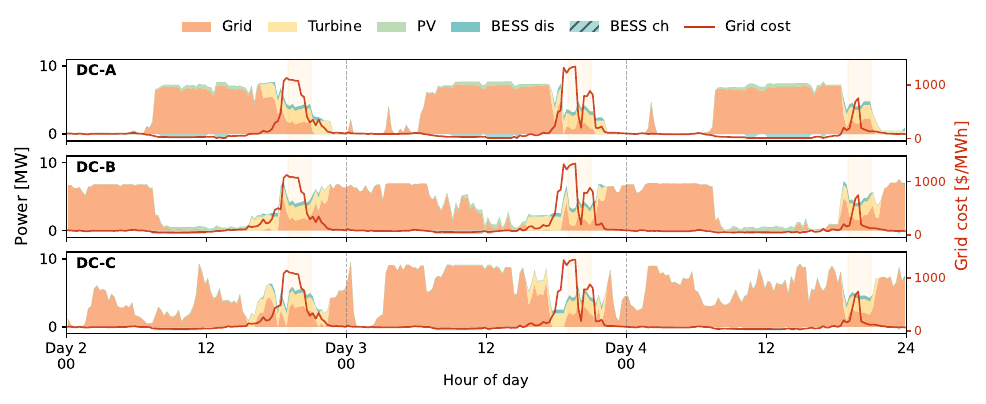}
\caption{\textcolor{black}{S6 power supply and grid costs over Days~2--4 (15-min resolution).}}
\label{fig:multiday_power_cost}
\end{figure}

\begin{figure}[!htbp]
\color{black}
\centering
\captionsetup{font=footnotesize,skip=2pt}
\includegraphics[width=\columnwidth]{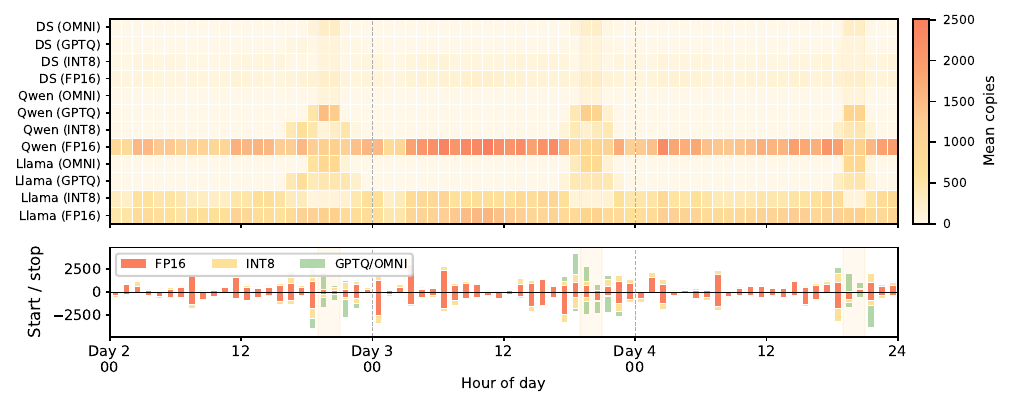}
\caption{\textcolor{black}{S6 loaded-instance counts (hourly means) and startup/shutdown counts (hourly sums) over Days~2--4.}}
\label{fig:multiday_instances}
\end{figure}

\begin{figure}[!htbp]
\color{black}
\centering
\captionsetup{font=footnotesize,skip=2pt}
\includegraphics[width=\columnwidth]{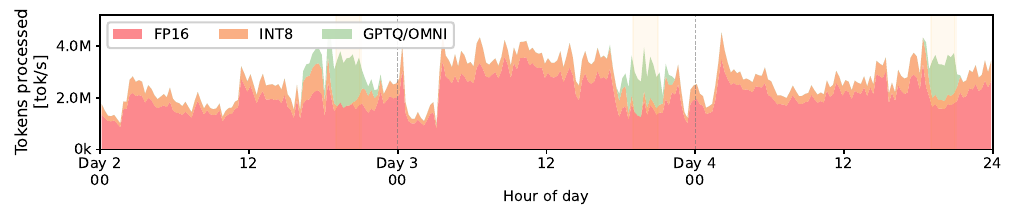}
\caption{\textcolor{black}{S6 token rates by precision over Days~2--4 (15-min resolution). In Figs.~\ref{fig:multiday_power_cost}--\ref{fig:multiday_token_rates}, dashed lines separate days and shaded bands mark the 19:00--21:00 DR windows. The underlying optimization retains 15-min intervals.}}
\label{fig:multiday_token_rates}
\end{figure}

\FloatBarrier 
\endgroup

\bibliographystyle{IEEEtran}
\bibliography{references_r1}


\begin{IEEEbiography}[{\includegraphics[width=1in,height=1.25in,clip,keepaspectratio]{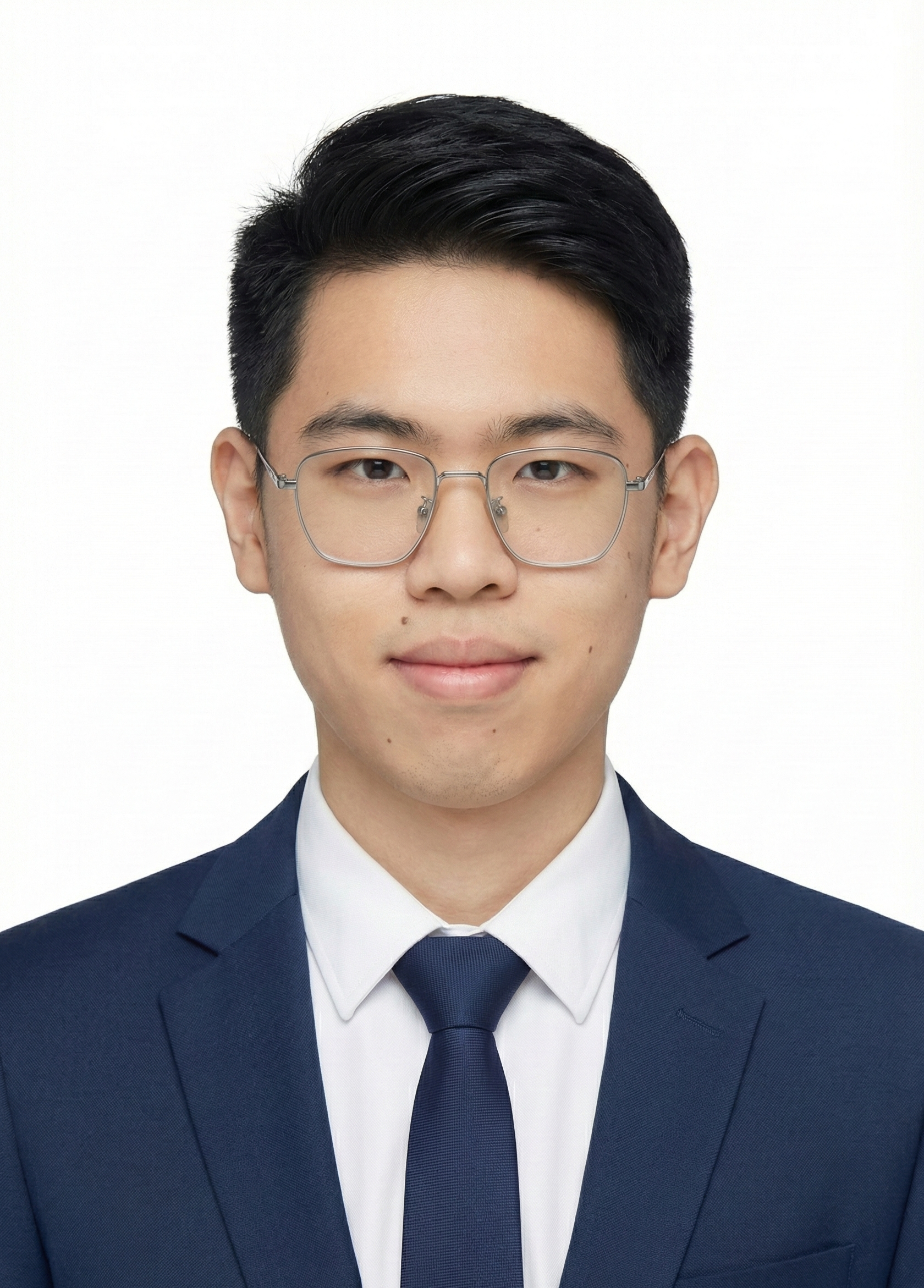}}]{Bojun Du (Student Member, IEEE)}
received the B.S. degree in fundamental science (mathematics and physics) and electrical engineering from Tsinghua University, Beijing, China. He is currently working toward the Ph.D. degree with the Department of Electrical Engineering, Tsinghua University. His research interests include computing--power coordination, large language models for power systems, power system foundation models, and constraint learning.
\end{IEEEbiography}

\begin{IEEEbiography}[{\includegraphics[width=1in,height=1.25in,clip,keepaspectratio]{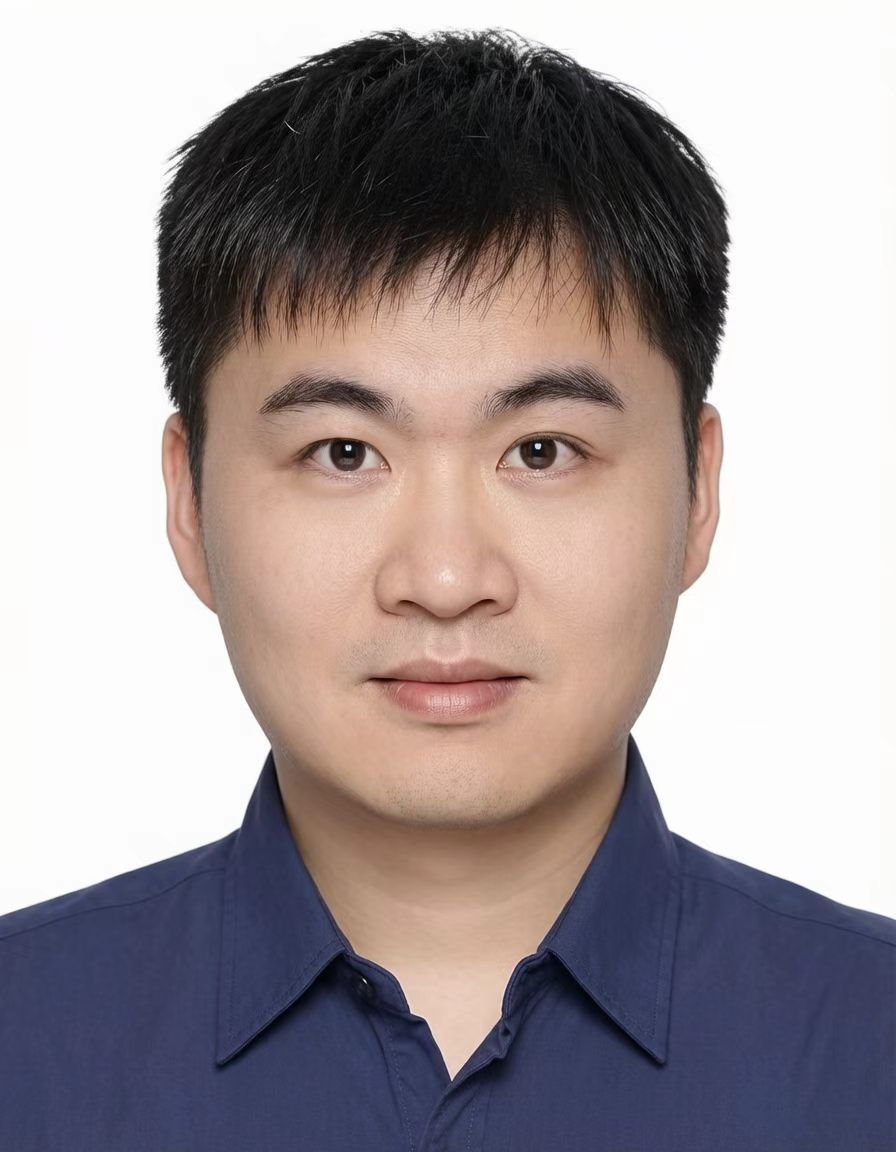}}]{Xiaoyi Fan (Member, IEEE)}
is currently a Full Professor at Shenzhen MSU-BIT University, Shenzhen, China. He received the Ph.D. degree from Simon Fraser University, Burnaby, Canada, in 2018, and the B.E. degree from the Beijing University of Posts and Telecommunications, Beijing, China, in 2013. He was a postdoctoral research fellow (honorary) with the Department of Electrical and Computer Engineering, University of British Columbia, Vancouver, Canada. He is a recipient of the NSERC Postdoctoral Fellowship in 2019. He is a recipient of the IEEE ICDCS Distinguished Paper Award (2024) and the IWQoS Best Poster Award (2024). His research interests include the Internet of Things, edge computing, and artificial intelligence.
\end{IEEEbiography}

\begin{IEEEbiography}[{\includegraphics[width=1in,height=1.25in,clip,keepaspectratio]{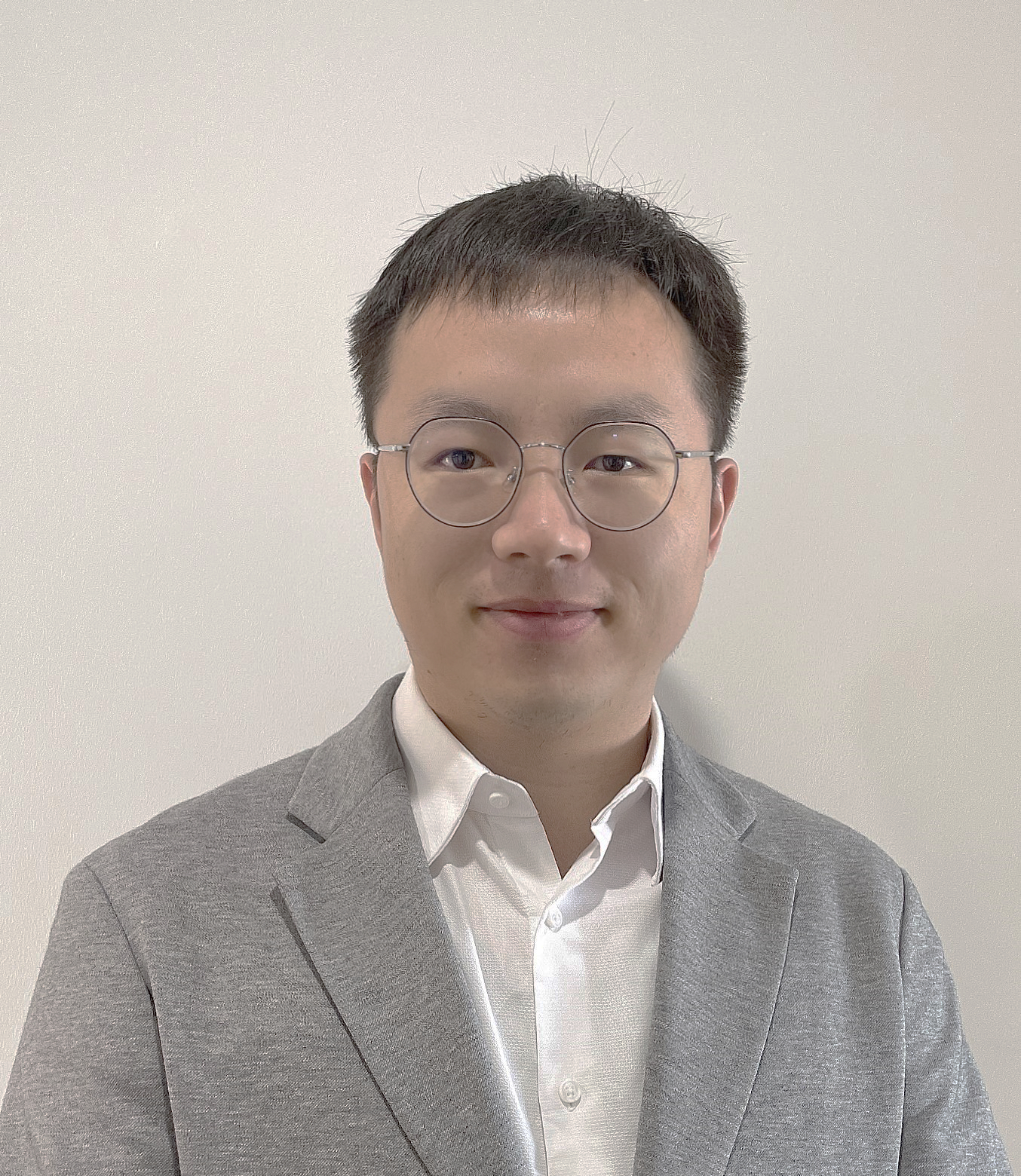}}]{Ershun Du (Member, IEEE)}
received the B.E. and Ph.D. degrees from the Department of Electrical Engineering, Tsinghua University in 2013 and 2018, respectively, where he is currently an Associate Professor. His research interests include low-carbon energy policy and climate change, renewable energy uncertainty analysis, and power system economics and planning.
\end{IEEEbiography}

\begin{IEEEbiography}[{\includegraphics[width=1in,height=1.25in,clip,keepaspectratio]{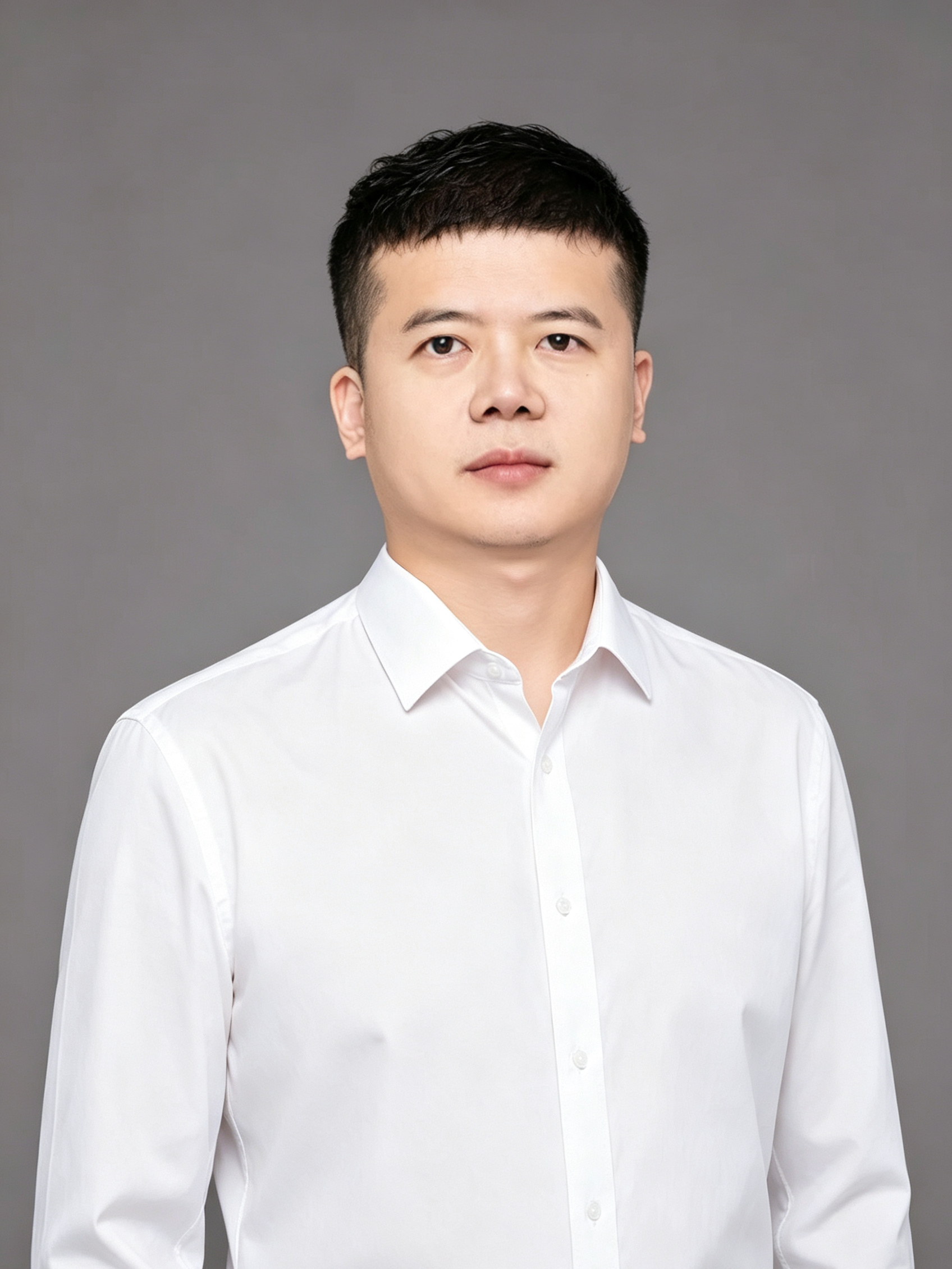}}]{Long Chen (Member, IEEE)}
received the B.E. and Ph.D. degrees in software engineering from Shanghai Jiao Tong University, China, in 2016 and 2022, respectively. He conducted postdoctoral research at Simon Fraser University, Canada, and The University of Hong Kong, Hong Kong. He previously worked as a Senior Researcher at ByteDance and is currently an Assistant Professor with The Hong Kong Polytechnic University. His research interests include energy-efficient AI systems, data-center energy management, and demand response in smart grids.
\end{IEEEbiography}

\begin{IEEEbiography}[{\includegraphics[width=1in,height=1.25in,clip,keepaspectratio]{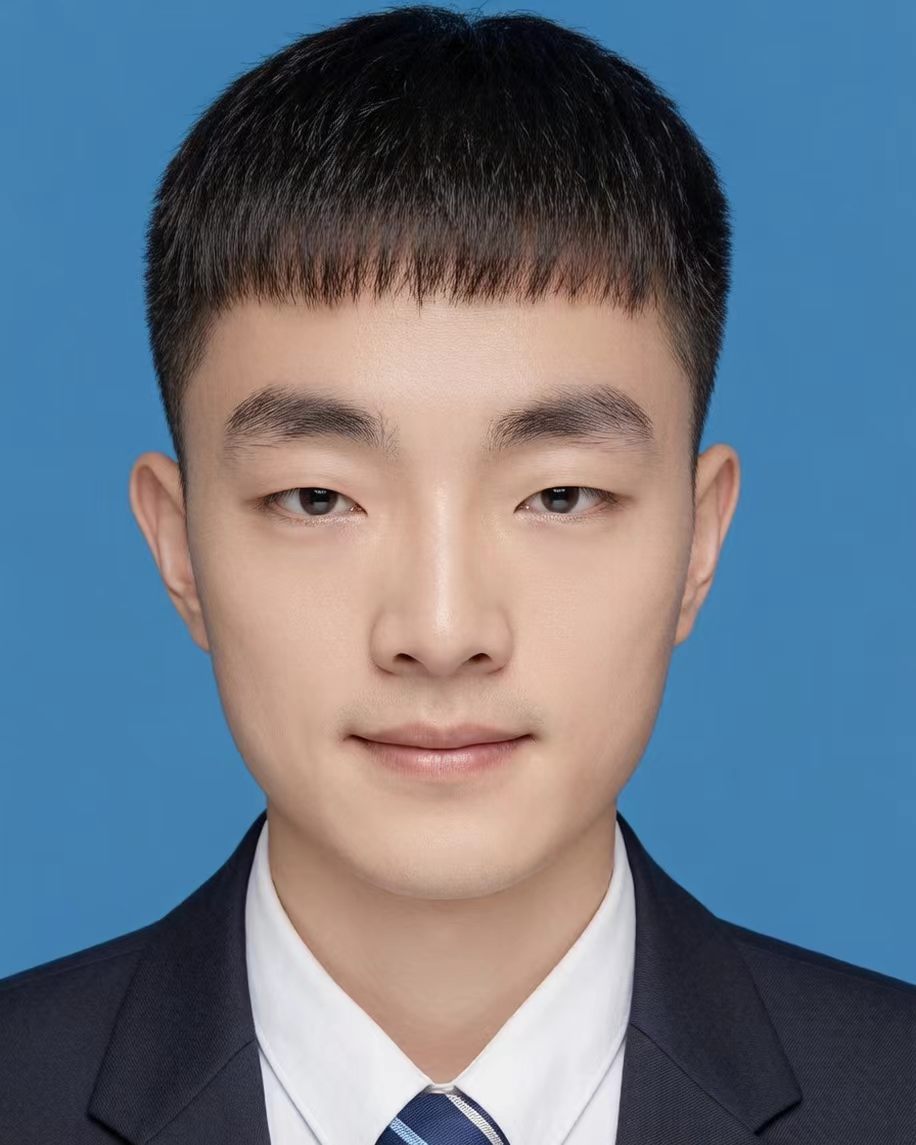}}]{Jianpei Han}
received the Ph.D. degree from the School of Electrical and Electronic Engineering, North China Electric Power University, Beijing, China, in 2023. From 2023 to 2025, he was a Postdoc Researcher with the Department of Electrical Engineering, Tsinghua University, Beijing, China. His research interests include power system optimization, demand response, collaboration of computing power, and electric power.
\end{IEEEbiography}

\begin{IEEEbiography}[{\includegraphics[width=1in,height=1.25in,clip,keepaspectratio]{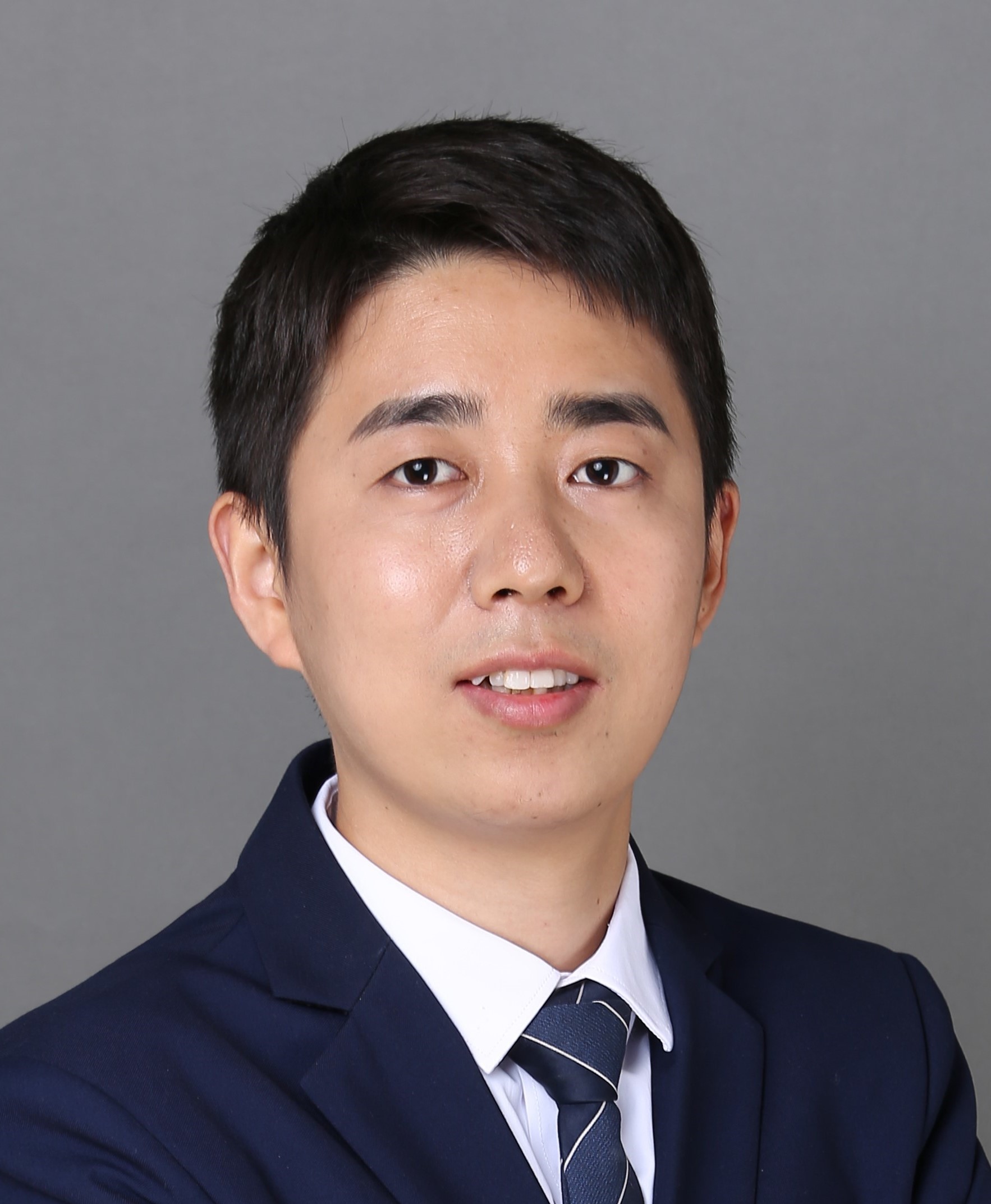}}]{Qingchun Hou (Member, IEEE)}
is an Assistant Professor and Researcher, and leads the AI-Driven Decision-Making and Optimization LAB (IDOLAB) at Zhejiang University. He received the Ph.D. degree from Tsinghua University in 2021, working with Prof. Chongqing Kang. He was a visiting Ph.D. student at the University of Washington from 2019 to 2020, working with Prof. Daniel Kirschen. He obtained the B.E. degree from the Huazhong University of Science and Technology in 2016. He worked as an Algorithm Expert with Alibaba Group from 2021 to 2024, leading a team to develop an AI-based combinatorial optimization solver (Laser) for digital supply chains and e-commerce platforms. His research interests encompass AI, LLM, RL, and optimization, along with their applications in power systems, smart grids, logistics, and supply chains. His current work centers around power system foundation models, data-driven power system operation and planning under high renewable energy penetration, AI for optimization solving and constraint satisfaction, and optimization for AI quantization. He was the recipient of the Advanced Technology Talent Program ``AliStar'' of Alibaba. His work has been published in top journals (Proceedings of the IEEE, IEEE Transactions on Power Systems, and IEEE Transactions on Smart Grid, etc.) and top AI conferences (ICLR, NeurIPS, AAAI, etc.).
\end{IEEEbiography}

\begin{IEEEbiography}[{\includegraphics[width=1in,height=1.25in,clip,keepaspectratio]{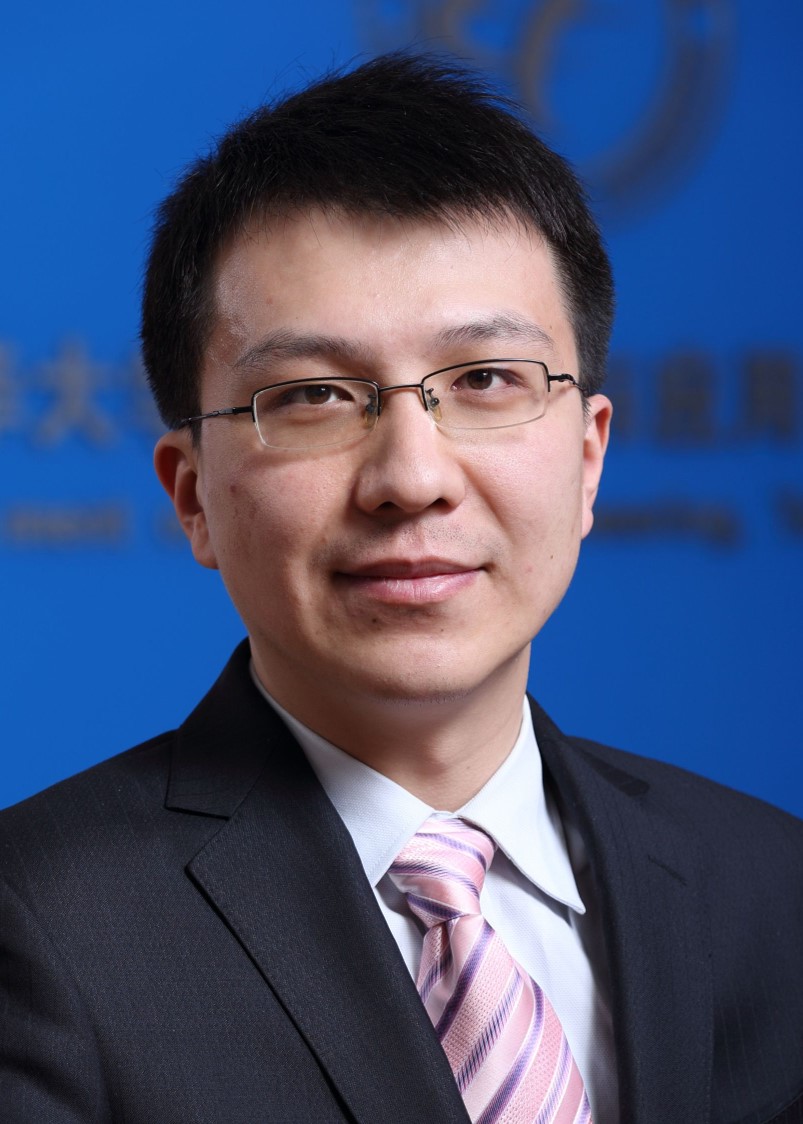}}]{Ning Zhang (Senior Member, IEEE)}
received the B.E. and Ph.D. degrees from Tsinghua University, Beijing, China, in 2007 and 2012, respectively. He is currently an Associate Professor with Tsinghua University. His research interests include power system planning and operation under renewable energy penetration, multiple energy systems integration, and stability-aware power system optimization.
\end{IEEEbiography}

\begin{IEEEbiography}[{\includegraphics[width=1in,height=1.25in,clip,keepaspectratio]{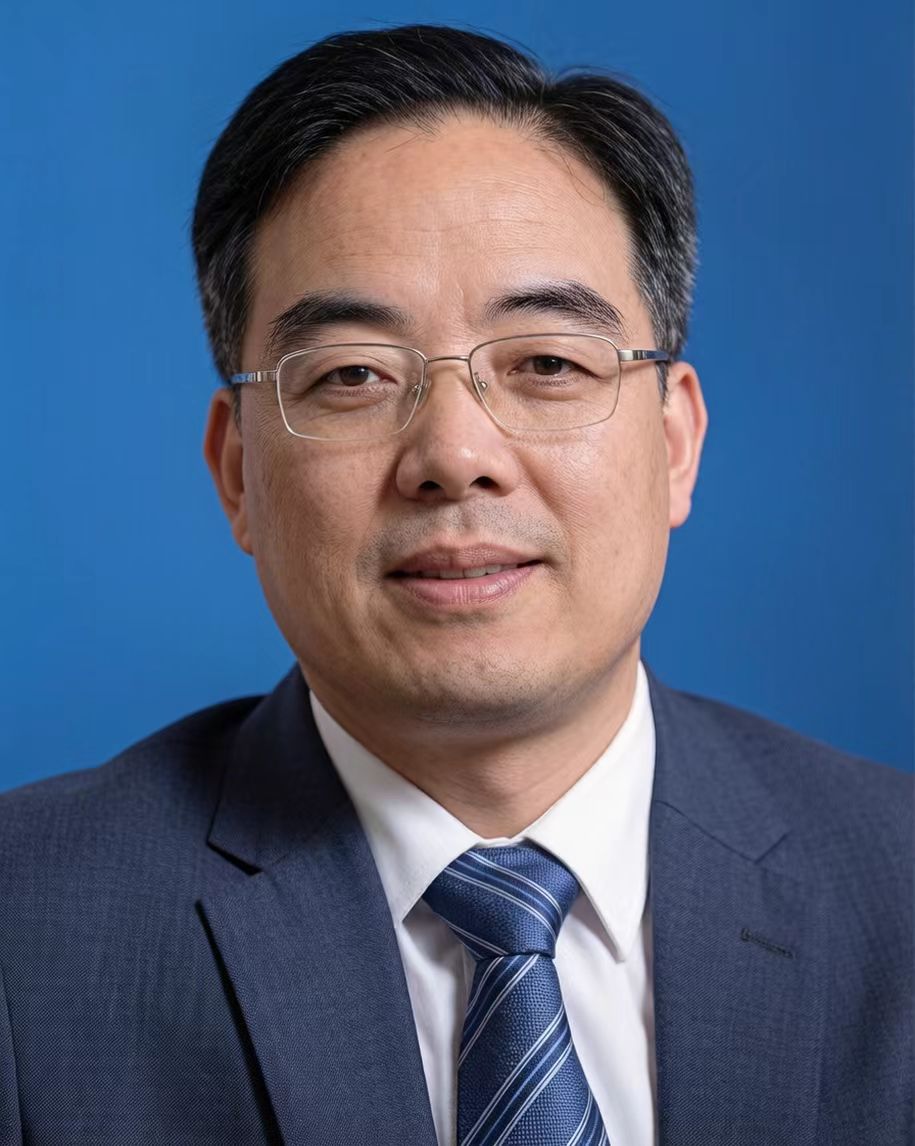}}]{Chongqing Kang (Fellow, IEEE)}
received the Ph.D. degree from the Department of Electrical Engineering, Tsinghua University, Beijing, China, in 1997, where he is currently a Professor. His research interests include power system planning, power system operation, renewable energy, low-carbon electricity technology, and load forecasting.
\end{IEEEbiography}

\end{document}